%% file: mff_big_paper_v5_arXiv.tex
\documentclass[twocolumn,trackchanges,floatfix]{aastex63}
\usepackage{amsmath}
\usepackage{hyperref}
\usepackage[T1]{fontenc}
\usepackage{xcolor}
\usepackage{soul}
\usepackage{subfigure}
\usepackage{graphicx}
\usepackage{todonotes}
\usepackage[version=4]{mhchem}
\usepackage{isotope}
\usepackage{multirow}
\usepackage{tablefootnote}
\usepackage{comment}
\usepackage{aas_macros}
\usepackage{graphicx}
\usepackage{dcolumn}
\usepackage{bm}
\usepackage{algorithm}
\usepackage{xspace}
\usepackage{algpseudocode}
\usepackage{appendix}
\usepackage{cancel}

\input{MFF_big_defs}

\newcommand{\imo}{\ensuremath{{\rm Im}(\Omega)}\xspace}
\newcommand{\kmax}{\ensuremath{|k|_{\rm max}}\xspace}
\definecolor{brickred}{rgb}{0.8, 0.25, 0.33}
\definecolor{darkgreen}{rgb}{0.0, 0.5, 0.0}

\newcommand{\revision}[1]{\textnormal{#1}}

\begin{document}

\reportnum{N3AS-23-029}

\title{Two-Moment Neutrino Flavor Transformation with applications to the Fast Flavor Instability in Neutron Star Mergers}

\author{Evan Grohs}
\affiliation{Department of Physics, North Carolina State University, Raleigh, NC 27695, USA}

\author{Sherwood Richers}
\affiliation{Department of Physics and Astronomy, University of Tennessee Knoxville, Knoxville, TN 37996, USA}

\author{Sean M. Couch}
\affiliation{Department of Physics and
Astronomy, Michigan State University,
East Lansing, Michigan 48824, USA}
\affiliation{Department of Computational Mathematics, Science, and
Engineering, Michigan State University,
East Lansing, Michigan 48824, USA}
\affiliation{Facility for Rare Isotope Beams,
Michigan State University,
East Lansing, Michigan 48824, USA}

\author{Francois Foucart}
\affiliation{Department of Physics \& Astronomy, University of New Hampshire, 9 Library Way, Durham NH 03824, USA}

\author{Julien Froustey}
\affiliation{Department of Physics, North Carolina State University, Raleigh, NC 27695, USA}
\affiliation{Department of Physics, University of California Berkeley, Berkeley, CA 94720, USA}

\author{James P. Kneller}
\affiliation{Department of Physics, North Carolina State University, Raleigh, NC 27695, USA}

\author{Gail McLaughlin}
\affiliation{Department of Physics, North Carolina State University, Raleigh, NC 27695, USA}

\begin{abstract}

Multi-messenger astrophysics has produced a wealth of data with much more to come in the future.  This enormous data set will reveal new insights into the physics of core collapse supernovae, neutron star mergers, and many other objects where it is actually possible, if not probable, that new physics is in operation. To tease out different possibilities, we will need to analyze signals from photons, neutrinos, gravitational waves, and chemical elements.  This task is made all the more difficult when it is necessary to evolve the neutrino component of the radiation field and associated quantum-mechanical property of flavor in order to model the astrophysical system of interest --- a numerical challenge that has not been addressed to this day. In this work, we take a step in this direction by adopting the technique of angular-integrated moments with a truncated tower of dynamical equations and a closure, convolving the flavor-transformation with spatial transport to evolve the neutrino radiation quantum field.
We show that moments capture the dynamical features of fast flavor instabilities in a variety of systems, although our technique is by no means a universal blueprint for solving fast flavor transformation. To evaluate the effectiveness of our moment results, we compare to a more precise particle-in-cell method.  Based on our results, we propose areas for improvement and application to complementary techniques in the future.

\end{abstract}


\section{Introduction}
\label{sec:intro}

Both neutron star mergers and core-collapse supernovae are true multi-messenger events, as they produce neutrinos, photons, gravitational waves, and chemical elements.  In the coming decade, there will be a wealth of data from all of these messengers, see e.g. \cite{Kalogera:2021bya,SNEWS:2021ewj,Holmbeck:2020kfc,Cowperthwaite:2017dyu,2009JCAP...01..047L,2009arXiv0912.0201L,bellm2018zwicky,tartaglia2018early}.  In order to produce the most realistic theoretical predictions to compare with future data, much theoretical development is still needed.  One significant area that requires attention is the neutrino physics of hot and dense systems (for a recent review see \citealt{Volpe:2023met}). 

Stellar explosions that reach extreme temperatures and densities, such as core-collapse supernovae and neutron star mergers, produce enough neutrinos that they account for a substantial portion of the energy budget (for recent estimates see \citealt{2021ApJ...915...28B,2019MNRAS.485.3153B,Fujibayashi:2022ftg,Hayashi:2021oxy,2023PhRvD.107j3055F,cusinato2022neutrino}). The majority of these neutrinos are in the energy range of tens of MeV.  In neutrino rich regions,  the ratio of neutrons to protons is influenced by electron neutrino and electron antineutrino capture reactions.  This neutron-to-proton ratio is a key factor influencing element synthesis (e.g., \citealt{McLaughlin:1996yzo,Freiburghaus:1999no,Surman:2005kf,Lippuner2015,2019ApJ...870....2C,Miller:2019mfl,2021MNRAS.501.5733R,2023ApJ...945L..13C}). 

Exploratory work has demonstrated the importance of accurately understanding the impact of changes in neutrino flavor as a function of both time and position in the exploding object. For example, the distribution of neutrinos among flavors influences: the outcome of the supernova explosions in one-dimensional high-mass CCSNe simulations \citep{Stapleford:2019yqg,Ehring:2023lcd,Ehring:2023abs}; and the results of element synthesis obtained in neutrino-cooled accretion disks (e.g., \citealt{Malkus:2012ts,Just:2022flt}), supernovae (e.g., \citealt{Duan:2010af, Mukhopadhyay:2022yrd,2023MNRAS.519.2623F}) and hypermassive neutron star outflows (e.g., \citealt{Fernandez:2022yyv,2020PhRvD.102j3015G,Li:2021vqj}).  Since the type of elements that are produced in ejecta depend sensitively on the ratio of neutrons to protons, these studies conclude that there is an impact on the elements that are produced in Core-Collapse SuperNovae (CCSNe) and Neutron Star Mergers (NSMs).

The Quantum Kinetic Equations (QKEs), where the terms representing the interactions of the neutrinos are expanded in a series, are often taken as a starting point for calculating the outcome of neutrino transport and propagation.
The first term in this series corresponds to evolution through a potential while the second term corresponds to momentum-changing collisions (e.g., \citealt{Volpe:2013uxl,2014PhRvD..89j5004V,2016PhRvD..94c3009B,Froustey:2020mcq}).
Different groups use the phrase QKE in a variety of manners, referring to the specific terms included/excluded in the series expansion.  In this work, we will use the phrase QKE to denote any equation which modifies the neutrino density matrices in time.
The starting point for classical neutrino transport can be obtained from the neutrino QKEs under the approximation that neutrino density matrices are always on-diagonal.  In this case, only the second term in the QKE series expansion -- the collision term -- is relevant.  

Modern codes aiming to perform global 3D simulations of CCSNe \citep{10.1093/mnras/stv1892,Kuroda_2016,Skinner_2019,Bruenn_2020} and/or NSMs and post-merger remnants \citep{Ruffert1999,Rosswog:2003rv,Wanajo2014,Neilsen:2014hha,Perego:2015agy,Foucart:2016rxm,Ardevol-Pulpillo:2018btx,Gizzi:2019awu,Foucart:2021mcb,Radice:2021jtw} that incorporate neutrinos generally use such classical transport algorithms. Given the difficulty of accurately solving the transport problem in global simulations with sufficient resolution and detailed microphysics, these codes inevitably need to make a range of additional approximations.
On the methods side, these might include the use of approximate transport schemes (leakage, truncated moments) or low-resolution Monte-Carlo methods, and/or the use of energy-integrated transport or ray-by-ray transport. On the microphysics side, this often includes the use of approximate interaction rates,  reduced number of neutrino flavors, or simply ignoring interactions that are too costly to calculate in practice, e.g., pair processes and inelastic scattering. While global neutrino transport algorithms are rapidly improving, they are still having significant difficulties in capturing all important aspects of the classical transport equation (e.g., \citet{2014ApJS..214...16N,2018ApJ...854..136N,2020ApJ...903...82I} for CCSNe and \citet{2019ApJS..241...30M,Miller:2019dpt} for NSMs).

Meanwhile, the starting point for flavor transformation in the absence of collisions is often studied by evolving the flavor field using only the first term in the QKE series expansion.  The evaluation of this term is done by use of operator splitting of the Hamiltonian (so-called \emph{mean-field}). This set-up has been studied extensively: for example,  the part of this Hamiltonian associated with neutrino coherent forward scattering on other neutrinos, in combination with other Hamiltonian terms, gives rise to the phenomenon of bipolar oscillations (for a review see \citealt{2006PhRvD..74l3004D}) and matter neutrino resonance transitions  \citep{Malkus:2012ts,Malkus:2014iqa,Frensel:2016fge,Wu:2015fga,Tian:2017xbr,2018PhRvD..97h3011V}. Additionally, Fast Flavor Conversion (FFC) stems from the combination of specific angular distributions of neutrinos, the mean-field Hamiltonian, and inclusion of neutrino advection, e.g. \cite{Sawyer:2005jk,Dasgupta:2016dbv,Izaguirre:2016gsx}. The relevant angular distributions are expected to occur in both supernovae, e.g. \citet{Abbar:2018shq, Nagakura:2021hyb,Nagakura:2023mhr}, and neutron star mergers \citep{Wu:2017qpc} at positions close to the central object. A number of works exist that evaluate classically computed angular distributions to determine whether these distributions have a Fast Flavor Instability (FFI), using a variety of techniques \citep{2018PhRvD..98j3001D,Johns:2021taz,2021PhRvD.104f3014N,Richers:2022dqa,Abbar:2023kta}.  The hallmark of a test for whether an instability will exist is to look for a ``crossing'' between a curve that represents the number density of neutrinos as a function of angle, and the curve that represents the number density of antineutrinos as a function of angle, e.g. \citet{Dasgupta:2009mg,Abbar:2017pkh,Dasgupta:2021gfs}.

Ideally one wishes to use both of the first two terms in the quantum kinetic equation series and some efforts have been undertaken with the inclusion of both. When including both the first and second term in the QKE series, the collision term most typically produces decoherence of the neutrinos, e.g. \citet{Richers2019}, and if that term is sufficiently large, the neutrinos tend to drift into flavor states.  However, under the right conditions, the combination of the two terms can also enhance flavor transformation through collisional instabilities \citep{Johns:2021qby,Johns:2022yqy,Xiong:2022vsy,Xiong:2022zqz}.

At present, there are questions about whether the QKEs can ever completely capture the behavior of neutrinos in these astrophysical systems, specifically because of the operator splitting technique that is used to write down the Hamiltonian.  There are ongoing efforts to analyze the evolution of neutrinos due to the forward scattering part of the many-body Hamiltonian under the assumption of continuous temporal interactions of all neutrinos with all other neutrinos \citep{Balantekin:2006tg,Pehlivan:2011hp,Siwach:2022xhx,Lacroix:2022krq,Balantekin:2023qvm,Cervia:2022pro, Rrapaj:2019pxz,2021PhRvD.104l3035P,Roggero:2022hpy,2023arXiv230107049M,2023arXiv230716793M}.

Notwithstanding these questions, efforts have been made to compute the QKEs by capturing the evolution of many neutrino ``packets'' in many different directions (e.g., \citealt{Sawyer:2005jk,2020PhLB..80035088M,2021PhRvD.104j3023R,PhysRevD.102.063018,2021PhRvD.104h3035Z,2022PhRvD.106f3011N,george_COSENuCollective_2022}). These methods provide useful benchmarks but are at present too computationally expensive to use extensively.  

An alternative is to use a reformulation of the QKEs in terms of the angular moments \citep{2013PhRvD..88j5009Z,Richers2019}. This reformulation creates a series of equations describing the time evolution of each moment and one then evolves only a small number of these equations. One then has to choose what to do with the moments which appear in the evolution equations but which are not explicitly evolved. One approach is simply to ignore the evolution of the moments above some order, however, what is found in practice is that one needs to retain a large number of the moment evolution equations making this approach computationally inefficient \citep{2018PhRvD..98j3001D, Johns:2019izj,2020PhRvD.102j3017J}.  An alternative solution to the truncation problem is to use a ``closure'' which links the unevolved moments to the lower order, evolved moments via some relationship. An example of calculations using this closure method can be found in \cite{Myers}, and using this approach, moment methods have been able to reproduce fast flavor transformations in neutron star merger-like conditions \citep{Grohs:2022fyq}.

In this manuscript, we consider in detail the efficacy of a two moment implementation of the QKEs neglecting the collision term and using a closure.
We illustrate our method using an example quantum closure that is a relatively straightforward extension of the classical maximum entropy closure.
In Sec.\ \ref{sec:math} we detail the QKE formalism and apply it to moments.
We elucidate the angular distributions corresponding to the closure and how they imply lepton number crossings in Sec.\ \ref{sec:mom_eln}.
Section \ref{sec:methods} gives an exposition of our implementation of neutrino flavor transformation in the framework of \flash \citep{2000ApJS..131..273F,DUBEY2009512} along with the initial and boundary conditions.
In Sec.\ \ref{sec:results}  we compare the results of our moment treatment to a more exact method for several well-studied test problems before turning to the presentation of results for three kinds of neutron star merger-like conditions.
We conclude and discuss the need for further exploration of moment QKE methods in Sec.\ \ref{sec:conclusion}.
With regard to units: we use two conventions.  When writing the neutrino flavor transformation equations, we use natural units where $\hbar=c=1$.  When giving results of numerical calculations, we use cgs units.

\section{The Moment Evolution Equations}
\label{sec:math}

\subsection{General Formalism}
\label{ssec:qke_gen}

We begin from the general QKEs describing the neutrino and anti-neutrino evolution adopting the nomenclature of \citet{1993NuPhB.406..423S,2014PhRvD..89j5004V,Froustey:2020mcq} and in particular \citet{2016PhRvD..94c3009B}. The evolved variable in the QKEs is a generalized density matrix for neutrinos, 
$\varrho = \varrho(t,\mathbf{x},\mathbf{p})$, and corresponding generalized density matrix $\varrhobar$ for anti-neutrinos, which are functions of time $t$, spatial location $\mathbf{x}$, and momentum $\mathbf{p}$.
In the treatment of \citet{2016PhRvD..94c3009B}, the generalized density matrices are one-body reduced density matrices~\citep{Volpe:2013uxl,Froustey:2020mcq}, and hereafter we will call $\varrho$ and $\varrhobar$ simply ``density matrices'' for the sake of brevity.
If we are describing neutrinos and anti-neutrinos with 2 chiral states, $\varrho$ and $\varrhobar$ are $2\,n_f\times2\,n_f$ Hermitian matrices for $n_f$ flavors. However in this work, we only consider left-chiral neutrinos and right-chiral anti-neutrinos, and ignore any kind of spin coherence \citep{2015PhLB..747...27C,2017PhRvD..95f3004T}.  As a result, the size of the density matrices is reduced to $n_f\times n_f$ for each of the neutrinos and anti-neutrinos.  In the case of three flavors of neutrinos, namely $e, \mu, \tau$, we can write the neutrino density matrix as the following
\begin{equation}\label{eq:dens_mat}
  \varrho = \begin{pmatrix}
  \varrho_{ee} & \varrho_{e\mu} & \varrho_{e\tau} \\
  \varrho_{\mu e} & \varrho_{\mu\mu} & \varrho_{\mu\tau} \\
  \varrho_{\tau e} & \varrho_{\tau\mu} & \varrho_{\tau\tau},
  \end{pmatrix}
\end{equation}
with a similar expression for anti-neutrinos.  In Eq.\ \eqref{eq:dens_mat}, the diagonal terms indicate the occupation numbers for a given flavor. The off-diagonal terms of Eq.\ \eqref{eq:dens_mat} encode the quantum coherence between two flavors.
The expressions for the number density, energy density, and the number flux vector are obtained by taking appropriate phase-space integrals of $\varrho$:
\begin{align}
  \mathcal{N}(t,\mathbf{x}) &= \frac{1}{(2\pi)^3}\int d^3p\, \varrho(t,\mathbf{x},\mathbf{p}) \, ,\label{eq:n_dens}\\
  \mathcal{E}(t,\mathbf{x}) &= \frac{1}{(2\pi)^3}\int d^3p\, p\,\varrho(t,\mathbf{x},\mathbf{p}) \, ,\label{eq:e_dens}\\
  \mathcal{F}^i(t,\mathbf{x}) &= \frac{1}{(2\pi)^3}\int d^3p\, \frac{p^i}{p}\,\varrho(t,\mathbf{x},\mathbf{p}) \, ,\label{eq:f_dens}
\end{align}
where the superscript $i$ indicates a component of a 3-vector. 
Note that we approximate neutrinos as ultrarelativistic by setting the neutrino energy equal to the 3-momentum magnitude $p$ and that $\mathcal{N}$, $\mathcal{E}$, and the components of $\mathbf{\mathcal{F}}$ are all $n_f \times n_f$ matrices.  The expressions for anti-neutrinos are analogous.

The QKEs for $\varrho$ and $\varrhobar$ are
\begin{align}
  \frac{\partial\varrho}{\partial t} + \dot{\mathbf{x}}\cdot\frac{\partial\varrho}{\partial\mathbf{x}} + \dot{\mathbf{p}}\cdot\frac{\partial\varrho}{\partial\mathbf{p}}
  &= -\imath\,[H,\varrho] + C,\label{eq:qke_nu}\\
  \frac{\partial\varrhobar}{\partial t} + \dot{\mathbf{x}}\cdot\frac{\partial\varrhobar}{\partial\mathbf{x}} + \dot{\mathbf{p}}\cdot\frac{\partial\varrhobar}{\partial\mathbf{p}}
  &= -\imath\,[\overline{H},\varrhobar] + \overline{C},\label{eq:qke_bnu}
\end{align}
where the single dot over a variable indicates differentiation with respect to time. In Eqs.\ \eqref{eq:qke_nu} and \eqref{eq:qke_bnu}, $C$ and $\overline{C}$ are collision terms which can change neutrino flavor, number, or momenta.  We shall ignore them throughout this work given the large separation of scales between the fast-flavor instability growth rate and the collision rates simulated here.  In addition, we will consider systems where the particle 3-momenta do not change with time, implying we may exclude the force term on the lhs of Eqs.\ \eqref{eq:qke_nu} and \eqref{eq:qke_bnu}.
To study flavor transformation, we employ Hamiltonian-like operators in Eqs.\ \eqref{eq:qke_nu} and \eqref{eq:qke_bnu} consistent with mean-field treatments.  When working to first order in power counting of the QKEs \citep{2014PhRvD..89j5004V}, the Hamiltonian operators are a sum of three potentials.  Specifically,
\begin{align}
  H &= H_V + H_M + H_{\nu},\\
  \overline{H} &= H_V - H_M - H_{\nu}^{\ast},\label{eq:bar_H}
\end{align}
denoting the vacuum $(H_V)$, matter $(H_M)$, and self-interaction $(H_{\nu})$ terms, and where ${}^{\ast}$ denotes complex conjugation.
The vacuum term arises from non-zero neutrino rest masses, and we write it as 
\begin{equation}
  H_V = \frac{1}{2p}\,U\,M^2\,U^\dagger \, ,
\end{equation}
where $U$ is the PMNS matrix and $M^2=\mathrm{diag}(m_1^2,m_2^2,m_3^2)$ is the diagonal matrix of squared neutrino masses.  The matter term is linear and familiar in the context of oscillations with solar neutrinos.  Electrons and positrons interact weakly with neutrinos in a flavor-dependent manner, which we denote by the following expression
in the case the matter fluid has zero velocity
\begin{equation}\label{eq:hm}
   H_M = \sqrt{2} G_F \, n_e \, I_e,
\end{equation}
where $G_F \simeq 1.166\times10^{-11}\,{\rm MeV}^{-2}$ is the Fermi coupling constant, $n_e$ is the difference between the number density of electrons and positrons, and $I_e$ is the electron flavor projection operator, i.e. $I_e={\rm diag}(1,0,0)$ for three flavors.  We will work in a frame comoving with the matter fluid implying Eq.\ \eqref{eq:hm} is valid (see App.\ \ref{sec:ortho}). 
Finally, the self-interaction potential is a consequence of neutrinos interacting with the background of other neutrinos
\begin{equation}
  H_{\nu} = \frac{\sqrt{2}G_F}{(2\pi)^3}\int d^3q(1-\cos\vartheta)[\varrho(t,\mathbf{x},\mathbf{q})-\varrhobar^{\ast}(t,\mathbf{x},\mathbf{q})],
  \label{eq:SI_hamiltonian}
\end{equation}
where $\vartheta$ is the angle between the free variable $\mathbf{p}$ and the integration variable $\mathbf{q}$.

\subsection{Moment Quantum Kinetic Equations}

In general, the density matrices are seven-dimensional since they depend upon time, space, and momentum. Solving the QKEs for the density matrices with sufficient temporal, spatial, and momentum resolution to ensure numerical convergence will be very computationally expensive. An alternative approach is to recast the QKEs as an infinite set of transport equations for the moments of the density matrices, and then truncate the number of moments that one solves by adopting a closure. Since moments are only five-dimensional quantities, solving their transport equations with sufficient fidelity to ensure convergence is a more feasible, though still difficult, computational challenge. 
In this paper we adopt a two moment scheme in which we evolve only the ``zeroth'' and ``first'' angular-integrated moments. 
We define the zeroth, first and second moment of $\varrho$ to be
\begin{align}
  E(t,\mathbf{x},p) &= \frac{p^3}{(2\pi)^3}\int d\Omega_p\, \varrho(t,\mathbf{x},\mathbf{p}),\label{eq:mom_0}\\
  F^i(t,\mathbf{x},p) &= \frac{p^3}{(2\pi)^3}\int d\Omega_p\frac{p^i}{p}\, \varrho(t,\mathbf{x},\mathbf{p}),\label{eq:mom_1}\\
  P^{ij}(t,\mathbf{x},p) &= \frac{p^3}{(2\pi)^3}\int d\Omega_p\frac{p^ip^j}{p^2}\, \varrho(t,\mathbf{x},\mathbf{p}),
\label{eq:mom_2}
\end{align}
where $i,j \in \{x,y,z\}$ are spatial indices. Note that a different convention was chosen with respect to \cite{Myers} (no $1/4\pi$ prefactors), for consistency with~\cite{Grohs:2022fyq}. Analogous expressions exist for the anti-neutrinos.
The integrals in these definitions are only over the momentum-space solid angle $\Omega_p$ i.e. the propagation directions of the neutrinos at a given spacetime location and comoving-frame neutrino energy, and not the entire phase-space as in Eqs.\ \eqref{eq:n_dens} -- \eqref{eq:f_dens}. 
Furthermore, the $p^3$ in the prefactor of Eqs.\ \eqref{eq:mom_0} -- \eqref{eq:mom_2} indicates that these are the differential energy density and differential energy flux. These are the quantities we have chosen to time-evolve in the \flash code since this is the convention used in many instances of classical neutrino moment transport. However we will oftentimes show results using instead the differential number density which is related to $E(t,\mathbf{x},p)$ via
\begin{equation}\label{eq:n_mom}
  N(t,\mathbf{x},p) = \frac{1}{p}\,E(t,\mathbf{x},p). 
\end{equation}
From the definition of the moments we see we can recover the expressions in Eqs.\ \eqref{eq:n_dens} -- \eqref{eq:f_dens} by integrating the moments over the neutrino energy $p$ i.e.
\begin{align}
    \mathcal{N}(t,\mathbf{x}) &= \int dp\, N(t,\mathbf{x},p) \, , \\
    \mathcal{E}(t,\mathbf{x}) &= \int dp\, E(t,\mathbf{x},p) \, ,\\
    \mathcal{F}^i(t,\mathbf{x}) &= \int dp\, \frac{F^i(t,\mathbf{x},p)}{p} \, .
    \label{eq:integrated_moments}
\end{align}
In this manuscript, we will only consider mono-energetic neutrinos, and as a result, our expressions for number density and number density moment differ by a factor of energy-bin width $\Delta p$.
Note that $F$ is the specific energy flux, but $\mathcal{F}$ is the energy-integrated number flux. For future reference, at this point we introduce the flux factor vector (actually a vector of matrices) which we define to be
\begin{equation}
  \mathbf{f}_{ab} = \frac{\mathbf{F}_{ab}}{E_{ab}},
\end{equation}
with the norm for a component of the flavor matrix defined as
\begin{equation}
  f_{ab}\equiv|\mathbf{f}_{ab}| = \sqrt{\sum\nolimits_i (f^i_{ab})^2},
\end{equation}
where $i$ runs over the spatial indices $x,y,z$.

From comparing the definitions of the moments and Eq.\ \eqref{eq:SI_hamiltonian} for self-interactions, we observe that the self-interaction term can be written as
\begin{equation}\label{eq:SI_moment}
  H_{\nu} = H_E - \frac{1}{p}\,\mathbf{p}\cdot \mathbf{H}_F,
\end{equation}
where the moment-self-interaction terms are
\begin{align}
  H_E &= \sqrt{2}G_F\left(\mathcal{N}-\overline{\mathcal{N}}^*\right),\label{eq:he}\\
  H_{F}^{j} &= \sqrt{2}G_F\left(\mathcal{F}^j-\overline{\mathcal{F}}^{*j}\right).\label{eq:hf}
\end{align}
We will use these moment self-interaction expressions for evolving our dependent variables of $E$ and $F^j$.
Note, however, that when writing Eqs.\ \eqref{eq:he} and \eqref{eq:hf}, the energy-integrated quantities $\mathcal{N}$ and $\mathcal{F}^j$ appear.  Indeed, for the particular physical phenomena we study here -- namely the FFI -- the ELN crossing depends on the number moments and not the energy ones ($\mathcal{E}$ and intensity).  The simulations we present in this work are for mono-energetic neutrino distributions where $N$ and $E$ are equal up to a units factor as shown in Eq.\ \eqref{eq:n_mom}. Nevertheless, we make the distinction between $\mathcal{N}$ and $\mathcal{E}$ for the FFI under the guise of an eventual incorporation of multi-energy distributions.

Although the physical systems which we model do occur in environments where general relativity has a pronounced effect over large distances, we will do all of our calculations in a local Minkowski reference frame where we may specify the spacetime metric as $g^{\mu\nu}={\rm diag}(-1,1,1,1)$. 
Therefore, the 3-vector contraction in Eq.\ \eqref{eq:SI_moment} (and all other subsequent 3-vector contractions in this work) is equivalent to the 3D dot product of Euclidean space. In addition, we will assume that gradients of the fluid velocity are locally approximately zero, which further simplifies the equations of motion.

With everything defined we can now write out the evolution equations for $E$ and $\mathbf{F}$ by performing moment integrations of Eqs.\ \eqref{eq:qke_nu} and \eqref{eq:qke_bnu} and scaling by appropriate factors of $p^3/(2\pi)^3$. In Cartesian coordinates the transport equations for the neutrino moments are 
\begin{align}
  \frac{\partial E}{\partial t} + \frac{\partial F^j}{\partial x^j}
  &= -\imath\,[H_V+H_M+H_E,E] + \imath\,[H_{F}^{j},F_j],\label{eq:qke_E}\\
  \frac{\partial F^j}{\partial t} + \frac{\partial P^{jk}}{\partial x^k}
  &= -\imath\,[H_V+H_M+H_E,F^j] + \imath\,[H_{F}^{k},P^{j}_{k}],\label{eq:qke_F}
\end{align}
where we have ignored the collision and force terms and assume the background matter to be homogeneous with zero velocity in the tetrad frame. Note that this form would also be applicable if we assume a Minkowski metric with nonzero velocity gradients for energy-integrated moments, but here we explicitly assume zero velocity everywhere.
Note also that the second (pressure) moment is not an evolved quantity in a M1 transport scheme and is calculated algebraically as a function of the two time-evolved moments $E$ and $\mathbf{F}$ using a closure relation. The closure relation we use will be discussed in Sec.\ \ref{ssec:mec}.  Finally for completeness, we give the moment evolution equations for the anti-neutrinos
\begin{align}
  \frac{\partial \overline{E}}{\partial t} + \frac{\partial \overline{F}^j}{\partial x^j}
  &= -\imath\,[H_V-H_M-H_E^{\ast},\overline{E}] +\imath\,[H_{F}^{j\ast},\overline{F}_j],\label{eq:qke_Ebar}\\
  \frac{\partial \overline{F}^j}{\partial t} + \frac{\partial \overline{P}^{jk}}{\partial x^k}
  &= -\imath\,[H_V-H_M-H_E^{\ast},\overline{F}^j] +\imath\,[H_{F}^{k\ast},\overline{P}^{j}_{k}].\label{eq:qke_Fbar}
\end{align}

Equations \eqref{eq:qke_E}, \eqref{eq:qke_F}, \eqref{eq:qke_Ebar}, and \eqref{eq:qke_Fbar} give the equations of motion for the neutrino field under study. They comprise a coupled set of matrix equations with spacetime indices $\{j,k\}$ ranging over the 3D space indices $\{1,2,3\}$. As an interesting aside, we note that it is possible to write the equations of motion in a 4D spacetime framework.  To begin, construct the following neutrino arrays
in the laboratory (Euler) frame
\begin{align}
  J^\alpha_{(\nu)} &= \begin{pmatrix}\mathcal{N} - \overline{\mathcal{N}}^*\\
  \boldsymbol{\mathcal{F}} - \overline{\boldsymbol{\mathcal{F}}}^*\end{pmatrix},\\
  T^{\alpha\beta}&=\begin{pmatrix}
  E & \mathbf{F} \\
  \mathbf{F}^{T} & \mathbf{P}
  \end{pmatrix},\label{eq:4d_t}\\
  H^\alpha &= -u^\alpha H_V - \sqrt{2} G_F(J^\alpha_{(e)}+J^\alpha_{(\nu)}), \label{eq:4d_h}\\
  \overline{H}^\alpha &= -u^\alpha H_V + \sqrt{2} G_F (J^\alpha_{(e)}+J^\alpha_{(\nu)})^*,\label{eq:4d_hbar}
\end{align}
where $J^{\alpha}_{(e)} = u^\alpha (n_e-\bar{n}_e)I_e$.
In the above Eq.\ \eqref{eq:4d_t}, $\mathbf{F}^{T}$ denotes the transpose of the row vector $\mathbf{F}$ into a column vector.
In addition, we define the four-velocity of the reference frame $u^\mu=(1,0,0,0)$ in Eqs.\ \eqref{eq:4d_h} and \eqref{eq:4d_hbar}.  With these definitions, we are able to cast the QKEs for neutrinos and anti-neutrinos as
\begin{align}
  \nabla_\beta T^{\alpha\beta} &= -i  \left[ H_\beta,T^{\alpha\beta}\right],\label{eq:four_eom_nu}\\
  \nabla_\beta \overline{T}^{\alpha\beta} &= -i  \left[\overline{H}_\beta,\overline{T}^{\alpha\beta}\right].\label{eq:four_eom_bnu}
\end{align}
In Eqs.\ \eqref{eq:four_eom_nu} and \eqref{eq:four_eom_bnu}, we adopt the convention where repeated indices are contracted with respect to the metric, i.e., $A^\alpha A_\alpha=g_{\alpha\beta}A^{\alpha}A^{\beta}$ with a $(-,+,+,+)$ convention.

\section{Moment Closure Relation and Lepton Number Crossing}
\label{sec:mom_eln}

\subsection{The Maximum Entropy Closure}
\label{ssec:mec}

The evolution equations for the fluxes $\mathbf{F}$ and $\overline{\mathbf{F}}$ involve the spatial gradients of the pressure tensors $P$ and $\overline{P}$; the evolution of the pressure tensors involve the spatial gradients of the next moment. This pattern continues in perpetuity and results in an infinite tower of equations. This is an unavoidable property of moment decomposition.
Nevertheless, in some situations the infinite set of equations can be solved: for example, when the radiation field is strongly-interacting, an equation of state will relate the pressure to the energy density under the assumption of Local Thermodynamic Equilibrium (LTE) thereby closing the set of equations for the first two moments. But in general -- and neutrinos in CCSNe and NSMs are both such cases -- no such equation of state exists that naturally closes the set of evolution equations. The simplest approach is to propose a local, analytic relation to close the tower of equations suited for the individual problem under study that matches analytic results in the trapped and free-streaming limits. This relation is called the closure relation (or ``closure'' for brevity). We will adopt this same approach when proposing a closure for the quantum moments of the neutrinos that must be able to account for both neutrino advection and the flavor transformation. We begin our explanation of the closure we adopt by ignoring the flavor structure of the moments for the time being, and consider the Maximum Entropy Closure (MEC) often used in ``classical'' moment transport. 

By definition of the MEC, the neutrinos of a particular species assume an angular distribution in momentum-space such that the angular entropy is extremized \citep{1978JQSRT..20..541M,1994ApJ...433..250C}. In other words, the neutrinos are distributed in the momentum-space angles such that an entropy-like function is maximized under the constraints of a net number density and flux.
These constraints relate directly to the dynamical variables of interest in Eqs.\ \eqref{eq:mom_0} and \eqref{eq:mom_1}. As with any reasonable closure, the MEC exactly represents the radiation field in the limits far from a source (where all radiation is moving in one direction) and when the radiation is in equilibrium. We will utilize the MEC for our flavor-mixing neutrino and anti-neutrino distributions, which we summarize below for completeness.

Under the constraints of number density and flux, the neutrino distribution of a particular species $a$ per unit solid angle of momentum-space, $\psi_{aa}$, is \citep{1978JQSRT..20..541M,1994ApJ...433..250C}
\begin{equation}\label{eq:dn_dOmega}
    \psi_{aa} = \frac{E_{aa}}{4\pi}\, \frac{Z_{aa}}{\mathrm{sinh}(Z_{aa})} e^{Z_{aa}\mu}\,\,,
\end{equation}
where $E_{aa}$ is the energy density moment for species $a$, and $\mu=\hat{F}_\mathrm{tet}\cdot\mathbf{\Omega}$ gives the angular dependence. 
$\mathbf{\Omega}$ is the direction unit vector and $\widehat{F}_{\rm tet}=\mathbf{F}_{aa}/F_{aa}$.
The parameter $Z_{aa}$ follows from the constraint on the magnitude of the flux factor vector $f_{aa}$ 
\begin{equation}
\begin{aligned}
    f_{aa}&=\frac{1}{E_{aa}}\int d\Omega\,\mu\,\psi_{aa}\\
    &=\coth Z_{aa} - \frac{1}{Z_{aa}}\,\,.
    \end{aligned}\label{eq:fn_z}
\end{equation}
Eq.\ \eqref{eq:fn_z} must, in general, be inverted numerically to obtain the value of $Z_{aa}$ corresponding to a given $f_{aa}$. Once $Z_{aa}$ is obtained, we can construct the angular distribution of the neutrinos in Eq.\ \eqref{eq:dn_dOmega}.
If we adopt initial neutrino distributions from a core collapse supernova or neutron star merger simulation that uses an MEC, then this is consistent with the assumptions in the original simulations. While the full angular information is assumed in our multi-direction calculations, our two-moment scheme simply uses the MEC to determine the pressure moment.  

Borrowing the terminology from classical radiation hydrodynamics, we interpolate the pressure moment between the optically thin and thick limits as
\begin{equation}
    P^{ij} = \frac{3\chi-1}{2}P^{ij}_\mathrm{thin}+\frac{3(1-\chi)}{2}P^{ij}_\mathrm{thick}\,\,,
    \label{eq:Pij_closed}
\end{equation}
where the thin and thick limits are
\begin{align}
  P_\mathrm{thin}^{ij} &= E \frac{F^i F^j}{F^2},\label{eq:P_thin}\\
  P_\mathrm{thick}^{ij} &= \frac{E}{3} \delta^{ij},\label{eq:P_thick}
\end{align}
and we have suppressed the flavor indices for ease in notation.
\citep{1978JQSRT..20..541M} demonstrate that these assumptions lead to a simple functional form of the Eddington factor $\chi$:
\begin{equation}
\chi = \frac{1}{3} + \frac{2}{15}f^2(3-f+3f^2),
\label{eq:ME_closure}
\end{equation}
such that Eq.\ \eqref{eq:Pij_closed} becomes consistent with the second angular moment of Eq.\ \eqref{eq:dn_dOmega}. In Sec.\ \ref{sec:BANG} we explore extending this concept to matrix-valued moments necessary for quantum neutrino transport.

\subsection{Lepton Number Crossing with the Maximum Entropy Closure}

At this point, we give a brief interlude to discuss lepton number crossings in the context of the MEC.
Assuming the momentum-space angular distributions of two neutrino species follow an MEC, one can analytically determine whether two distributions cross \citep{Johns:2021taz,Richers:2022dqa}. Such crossings (most straightforwardly between electron neutrino and anti-neutrino distributions) herald flavor instabilities \citep{morinaga2022fast}.
Although the conditions for neutrino flavor instability are more general and involve the other flavor-lepton-numbers, we shall consider initial conditions where the $x$-flavor Lepton Number (XLN) is zero and therefore the ELN crossing is the sole source of the instability.
We note that although only an ELN crossing is initially present, an XLN crossing can subsequently appear during the evolution.

The intersection of the angular distributions is the boundary of a 2D surface in the 3D momentum-space.  Solving for the boundary is, in general a difficult problem.  However, for the purposes of FFC, simply identifying whether or not the intersection exists suffices to determine whether the system is unstable or not.  Therefore, we can look at the 2D cross-sectional slice of the 3D distributions in the plane of both flux vectors to determine whether there is a ELN crossing or not.

We will determine whether an ELN crossing exists using energy density distributions, yet an ELN crossing utilizes number density distributions by definition.  However, we stress that our energy and number variables are simply related by a constant of proportionality for mono-energetic distributions, and as a result, we will continue using the energy quantities below.
Let $E_{ee}$ and $Z_{ee}$ define the maximum entropy distributions for electron neutrinos, and similarly $\overline{E}_{ee}$ and $\overline{Z}_{ee}$ for electron anti-neutrinos [see Eqs.~\eqref{eq:dn_dOmega} and~\eqref{eq:fn_z}]. Furthermore, we assume that the flux factors are separated by an angle $\theta$, i.e.,
\begin{equation}
  \cos\theta = \frac{\mathbf{f}_{ee}\cdot\overline{\mathbf{f}}_{ee}}{f_{ee}\overline{f}_{ee}}.
\end{equation}
The distributions cross if \citep{Richers:2022dqa}
\begin{equation}\label{eq:eln_cross}
\frac{\eta^2}{\alpha^2 + \gamma^2} \leq 1\,\,,
\end{equation}
where
\begin{align}
  \eta &= \ln\left[\frac{E_{ee} Z_{ee} \sinh(\overline{Z}_{ee})}{\overline{E}_{ee} \overline{Z}_{ee} \sinh(Z_{ee})}\right],\\
  \alpha &= \bar{Z}_{ee} \sin \theta,\\
  \gamma &= \bar{Z}_{ee} \cos \theta - Z_{ee}.
\end{align}
We use the criterion in Eq.\ \eqref{eq:eln_cross} to indicate the presence of FFI when choosing the locations from NSM simulations to consider in Sec.\ \ref{sec:results_nsm}.

\section{Methods}
\label{sec:methods}

We have written four QKEs (one energy density and three components for the flux density) in Eqs.\ \eqref{eq:qke_E} and \eqref{eq:qke_F}.  Along with the equations for the anti-neutrinos, this set of coupled matrix equations comprises 32 evolution variables per energy bin per spatial cell.  Our goal will be to integrate these equations under the conditions of FFI to see if this method can capture the behavior of FFC.  Before presenting results of test and NSM simulations, we give some more of the pertinent details on the numerical implementation of the moment method into \flash.  In addition, we give a brief exposition on the Particle-In-Cell (PIC) method in \emu and how it was tailored to compare with \flash.

\subsection{\flash}
\label{sec:BANG}

To study neutrino flavor transformation with moments, we use the \flash radiation hydrodynamics code \citep{2000ApJS..131..273F,DUBEY2009512},
further modified by \citet{2018ApJ...854...63O} which includes an M1 moment scheme for classical neutrino transport. 
It evolves the energy and flux density moments for three species: $\nu_e$, $\overline{\nu}_e$, and $\nu_\mathrm{other}$ for all other neutrinos. 
We modify the classical code by by distinguishing ${\overline{\nu}}_\mathrm{other}$ from $\nu_\mathrm{other}$, and adding the flavor off-diagonal components of the moments. This yields a total of eight effective species that follow from the generalized density matrices of Sec.\ \ref{ssec:qke_gen}, and specifically a 2-flavor version of Eq.\ \eqref{eq:dens_mat}. 
For example, these eight species for the energy density moments are $E_{ee}$, $E_{\mu\mu}$, $\mathrm{Re}(E_{e\mu})$, and $\mathrm{Im}(E_{e\mu})$ and the four charge-conjugate counterparts for the anti-neutrinos.  
In reality, neutrinos would oscillate between $e$ and the other two flavors, namely, $\mu$ and $\tau$.  Our implementation of flavor mixing is 2-flavor for simplicity, 
which artificially assumes that half of the heavy-lepton neutrinos and antineutrinos do not participate in flavor conversion,
but the number of flavors does not alter our qualitative conclusions and it will be possible to implement an 18-species framework for 3-flavors in the future. 

We decompose the domain into cells and group cells together into blocks to parallelize the computation over processors. Each block contains $16^3$ cells along with ghost cells.  The choice of $16^3$ cells per block results, in part, from the computational resources we use for this work.  For a different platform, we would be free to change the size of the blocks depending on the number of cores and available memory.
We use 6 ghost cells in each dimension so communication occurs only at the end of each full timestep, given a stencil size of 2 in each direction and 3 substeps within each full step\footnote{The stencil size is the number of grid cells referenced when evaluating a numerical derivative}.
The three-step integrator was originally designed to ensure consistency in the hydrodynamic evolution in \flash using a general tabulated equation of state. Our calculations do not evolve the hydrodynamics and add an unnecessary computational cost, but we leave the structure in place to ensure future consistency with the full \flash framework used for ab-initio compact object simulations.

We extend the 3-species transport subroutines from Appendix B of \citet{2018ApJ...854...63O} to the 8 species needed for flavor mixing. We use the same Harten-Lax-van Leer-Einfeldt (HLLE) Riemann solver \citep{harten1983upstream} to compute fluxes between grid cells for all 8 species.  We use a first-order method to reconstruct the interface flux and pressure values, instead of the second order TVD reconstruction employed in \citet{2018ApJ...854...63O}. Our advection timestep is set to $0.4$ times the grid cell light crossing time.

The advection and mixing evolution is done using an operator-split method, where the mixing derivatives are given by the rhs of Eqs.\ \eqref{eq:qke_E}, \eqref{eq:qke_F}, \eqref{eq:qke_Ebar}, and \eqref{eq:qke_Fbar}.  To calculate the commutators of $2\times2$ matrices, we decompose the density matrix into components as detailed above and use the commutation relations of the Pauli matrices.  Mixing is only treated locally, with the Hamiltonian-like terms specified at a given $\mathbf{x}$, or equivalently a given cell.  Unlike the solver for the advection, we use an adaptive, explicit 5th order Runge Kutta Cash-Karp (RKCK) method \citep{Press:1993:NRF:563041} in the mixing subroutine closely following the implementation in \citet{2016PhRvD..93h3522G}. 
The timestep is determined by requiring that the difference between the embedded 4th and 5th order solutions is smaller than one part in $10^6$ for each timestep and violations in unitarity (i.e., particle number conservation) are smaller than one part in $10^3$.

Finally, we discuss the MEC as implemented for mixing. The Eddington factor in Eq.\ \eqref{eq:ME_closure} is a derived result from the assumptions of a classical distribution maximizing angular entropy. Although our evolved quantities are expressed in a particular flavor basis, the physical evolution should not be basis dependent. Naively evaluating flux factors and Eddington factors using Eq.\ \eqref{eq:ME_closure} would break basis independence.  We could diagonalize the energy density moment such that the off-diagonal components of $E$ are zero.  We would also need to apply the same unitary transformation to each vector component of $\mathbf{F}$, but there is no guarantee that $E$ and $F^i$ are simultaneously diagonalizable. 
In addition, the flux factors of the flavor off-diagonal quantities are in general complex and can be arbitrarily large or small irrespective of whether the radiation is in the trapped or free-streaming regime, making naive flux factors for flavor off-diagonal components a poor choice for interpolating between these regimes.

To ameliorate these issues, we can make the assumed pressure tensor independent of the flavor basis if we calculate a single $\chi$ for neutrinos and a single $\overline{\chi}$ for antineutrinos using flavor-traced flux factors. Specifically, those flavor-traced flux factors are defined as
\begin{align}
f^{(FT)} &= \frac{|{\rm Tr}[\mathbf{F}]|}{{\rm Tr}[E]}\\
&= \frac{|(F^i_{ee} + F^i_{xx})\hat{x}_i|}{E_{ee} + E_{xx}},\label{eq:f_FT}
\end{align}
and a similar expression for the anti-neutrinos and $\overline{f}^{(FT)}$, where $\hat{x}_i$ are the Cartesian unit vectors.
These flavor traced flux factors are substituted into Eq.\ \eqref{eq:ME_closure} to obtain $\chi$ and $\bar{\chi}$, which are in turn used for all flavor components in Eq.\ \eqref{eq:Pij_closed}. This also prevents the flavor off-diagonal components from appearing to be in the optically thick regime when the flavor off-diagonal components are in the free-streaming regime. Note however, the principal direction of the pressure tensor is computed as in Eq.\ \eqref{eq:Pij_closed} separately for \emph{each} flavor component, i.e., $\varrho_{ee}$, $\varrho_{xx}$, ${\rm Re}[\varrho_{ex}]$, and ${\rm Im}[\varrho_{ex}]$.  In other words, Eqs.\ \eqref{eq:Pij_closed}, \eqref{eq:P_thin}, and \eqref{eq:P_thick} all have flavor indices on each quantity, except for $\chi$.

This scheme has the disadvantage that in the limit of no flavor mixing, it does not reduce to the original two moment transport scheme, since different flavors are no longer allowed to have independent flux factors. However, for many of the cases we study in Sec.\ \ref{sec:results}, we are in the optically thick limit for both $e$ and $x$ species, and as a result $f_{ee}\sim f_{xx}\sim f^{(FT)}$ and similarly for the anti-neutrino flux factors.  We leave a more detailed analysis of possible closures to future work \citep{kneller23}.

\subsection{\emu}
\label{sec:emu}

\emu \citep{2021PhRvD.103h3013R} is a three-dimensional particle-in-cell neutrino flavor transformation code that evolves Eqs.\ \eqref{eq:qke_nu}-\eqref{eq:qke_bnu} individually for a large number of computational particles. To evaluate the self-interaction part of the Hamiltonian [Eq.\ \eqref{eq:SI_hamiltonian}] we collect the contributions of each particle to the background angular moments of the distribution using a second order shape function, and interpolate the Hamiltonian from the grid to each particle using the same second-order shape function. The advection terms are accounted for by simply translating the position of each computational particle. The flavor density matrix and positions of each computational particle are evolved with a global fourth-order Runge-Kutta method. The snippets of \emu code that depend on the number of neutrino flavors are automatically generated using {\tt sympy} \citep{sympy} to carry out symbolic matrix operations, simplify the expressions, and output C++ code. This allows us to run simulations assuming either two or three neutrino flavors. \emu is publicly available at \cite{don_e_willcox_2021_5514018}.

Whereas \flash is a moment method and only evolves two angular moments for each flavor component of the neutrino distribution, \emu simulates particles moving in many individual directions. The \emu results we present in this paper were computed to 378 particles per cell corresponding to an angular resolution of roughly 11 degrees, following the resolution tests in \cite{2021PhRvD.104j3023R}.

\subsection{Initial and Boundary Conditions}
\label{subsec:initial_boundary_conditions}

We assign the flavor-diagonal values to the first two moments at every point in the domain for a \flash calculation.  There are eight neutrino species with four values (one energy density and three flux components), for a total of 32 initial values per cell which we need to assign.  However, in practice we always begin with identical moments for $\varrho_{xx}$ and $\overline{\varrho}_{xx}$, since all heavy lepton neutrino and antineutrino species are gathered in a single species $\nu_\mathrm{other}$ in \cite{Foucart:2016rxm} due to their very similar evolution in the absence of flavor transformation.

Our calculations of the FFI will only include the self-interacting term for the Hamiltonian-like operator in Eqs.\ \eqref{eq:qke_E}, \eqref{eq:qke_F}, and the anti-neutrino counterparts.  We set the vacuum and matter potentials to zero so as to focus on the FFI, leaving the interesting physics cases of slow collective modes and matter-neutrino resonances to future work.  In an actual astrophysical object, such as a CCSN or NSM, the vacuum potential would act to seed the flavor off-diagonal elements as a function of path length and neutrino energy.  Since we simulate a local volume within a larger global system and thus have no information about the advection of perturbed neutrinos into our domain, we choose to take precise manual control over the initial seeds and exclude the vacuum potential. We seed the off-diagonal flavor components with a perturbation of $\mathcal{O}(10^{-6})$ compared to the diagonal components.  The scale $10^{-6}$ is chosen such that the growth in the off-diagonal components begins in the linear regime. We have verified that starting with even smaller perturbations does not change the outcome for \flash calculations. This is expected, as the growth should be the same in the linear regime, and thus smaller initial perturbations only take longer and use more computing resources.  For the pattern of the perturbations, we use random numbers in each cell in order to remain agnostic to the scale of the initial perturbations that would be present in nature.

Specifically, we use the following to seed the initial perturbations in the off-diagonal components of the energy densities in \flash
\begin{equation}\label{eq:od_seed}
  \delta E_{ab}(\mathbf{x}) 
  = 10^{-6}\,p\,\max_c\{N_{cc}\}[A_{ab}(\mathbf{x})+iB_{ab}(\mathbf{x})],
\end{equation}
where $-1<A,B<1$ are uniform random numbers at each location $\mathbf{x}$, $N_{cc}$ are the initial number density moments for the diagonal components, and the Hermitian perturbation is only applied for $a\ne b$.
For the flux moment, we copy the energy density moment perturbation into the flux moment and use flux factor vectors to weight the direction
\begin{equation}\label{eq:flux_od_seed}
  \delta\mathbf{F}_{ab}(\mathbf{x}) = \delta E_{ab}(\mathbf{x})\frac{\Sigma_c N_{cc}\mathbf{f}_{cc}}{\Sigma_c N_{cc}},
\end{equation}
implying the initial perturbations for the off-diagonal components of $E$ and $\mathbf{F}$ are correlated.  Analogous expressions exists for the anti-neutrino moments.

In the \emu calculations, each particle is assigned a 4-momentum vector, weight, and density matrix.  The 4-momentum vectors are distributed uniformly in space, but assigned initial weights and density matrices to approximate the maximum entropy distribution [Eq.\ \eqref{eq:dn_dOmega}] separately for each flavor. In this way, the zeroth and first moments are reproduced under an appropriate angular integration for each flavor-diagonal element of the density matrix. We impose a random perturbation to the flavor off-diagonal elements of the density matrix at the level of $10^{-6}$ and adjust the diagonal values accordingly to preserve the length of each polarization vector.

The random numbers are determined at run time, so although the bulk properties of the instability are expected to be related by the similar initial conditions, the exact values in a particular cell have no correspondence between \emu and \flash calculations. We will make all comparisons in the aggregate between the two sets of calculations.
Furthermore, if we calculate the energy density for \emu using Eq.\ \eqref{eq:mom_0} (where the angular integration becomes a sum over particle index), we would expect the incoherent sum for $\delta E_{ab}$ to be reduced by $\sqrt{n}$ for $n$ particles per cell, implying an effectively smaller perturbation on the initial moments. To reiterate, this only impacts the length of the linear phase of the instability -- not the growth rate or saturation properties.
\revision{We give more details on the differences in the initial conditions between \flash and \emu in App.\ \ref{app:diff}.}

In both \flash and \emu calculations, we use a 3D cubic box with Cartesian coordinates.  The domain sizes and resolutions of the \flash production simulations are listed in Tables~\ref{tab:3test_parameters} and \ref{tab:NSM_parameters}. 
We choose the domain size and cell count so that we have the resolution to resolve the fastest growing mode in the FFI, along with enough of a spatial domain to contain a few wavelengths of that fastest growing mode. We do not know the properties of the fastest growing mode a priori, so we perform convergence checks inline with the presentation of the results.  The simulation durations are generally longer than the light crossing time of the domain, implying the initial particles/densities will have advected out of the domain before the end of the simulation.  We implement periodic boundary conditions for both sets of calculations implicitly assuming that the initial distribution is reasonably approximated as periodic on scales larger than the simulation domain. We also verify that changing the domain size does not impact the results.

\section{Results}
\label{sec:results}

Our results comparing the ability of the two-moment method to reproduce the fast-flavor instability are split into two parts. First, we consider in Section~\ref{sec:results_tests} the three test problems in 3-dimensions that were previously studied with \emu in \citet{2021PhRvD.104j3023R}. 
In Section~\ref{sec:results_nsm} we move to consideration of conditions extracted from a dynamical neutron star merger simulation.  We use the symbol \imo to denote the growth rate of $|N_{ex}|$ during instability.  In addition, we use the symbol \kmax to denote the fastest growing mode in the discrete Fourier transform of $N_{ex}$ during instability.

\subsection{3D Test Problems}
\label{sec:results_tests}

\begin{table*}
    \centering
    \hspace{-2.5cm}
    \begin{tabular}{c|cccccccc}
        \multirow{2}{*}{Name} & $\mathcal{N}_{ee}$ & $\overline{\mathcal{N}}_{ee}$ & $\Sigma \mathcal{N}_{(x)}$ & $\mathbf{f}_{ee}$ & $\overline{\mathbf{f}}_{ee}$ & $\mathbf{f}_{(x)}$ & $L$ & $N_{gp}$\tabularnewline
         & $(10^{32}\,\mathrm{cm}^{-3})$ & $(10^{32}\,\mathrm{cm}^{-3})$ & $(10^{32}\,\mathrm{cm}^{-3})$ & &&& (cm)\tabularnewline \hline
        Fiducial & $4.89$ & $4.89$ & 0 & $(0, \quad   0 \quad \ ,\ 1/3 \quad )$ & $(0, \quad   0 \quad \  , \ \, -1/3 \ \ )$ & $(0,0,0)$ & 8 & $128^3$\tabularnewline
        90Degree & $4.89$ & $4.89$ & 0 & $(0,1/\sqrt{18},1/\sqrt{18})$ & $(0,1/\sqrt{18},-1/\sqrt{18})$ & $(0,0,0)$ & 8 & $128^3$\tabularnewline
        TwoThirds & $4.89$ & $3.26$ & 0 & $(0,\quad   0 \quad \  , \quad   0 \quad \ )$ & $(0, \quad   0 \quad \ , \ \, -1/3 \ \ )$ & $(0,0,0)$ & 32 & $128^3$
    \end{tabular}
    \caption{List of simulation parameters and initial conditions for the three 3D test simulations. The first three columns show the number densities of each anti/neutrino flavor. For clarity, the third column shows the sum of all four heavy lepton anti/neutrino densities. Three-flavor simulations assume $\mathcal{N}_{\mu\mu}=\overline{\mathcal{N}}_{\mu\mu}=\mathcal{N}_{\tau\tau}=\overline{\mathcal{N}}_{\tau\tau}=\Sigma \mathcal{N}_{(x)}/4$, while two-flavor simulations assume $\mathcal{N}_{xx}=\overline{\mathcal{N}}_{xx}=\Sigma \mathcal{N}_{(x)}/4$, where we assume that the other half of the heavy-lepton neutrinos do not participate in flavor mixing. The next three columns show the flux factor vectors, the norm of which are the flux factors. The seventh column shows the length of each side of the domain and the eighth column the number of grid points for the baseline simulation.
    }
    \label{tab:3test_parameters}
\end{table*}

\begin{figure*}[htb!]
    \centering
    \includegraphics[width=\linewidth]{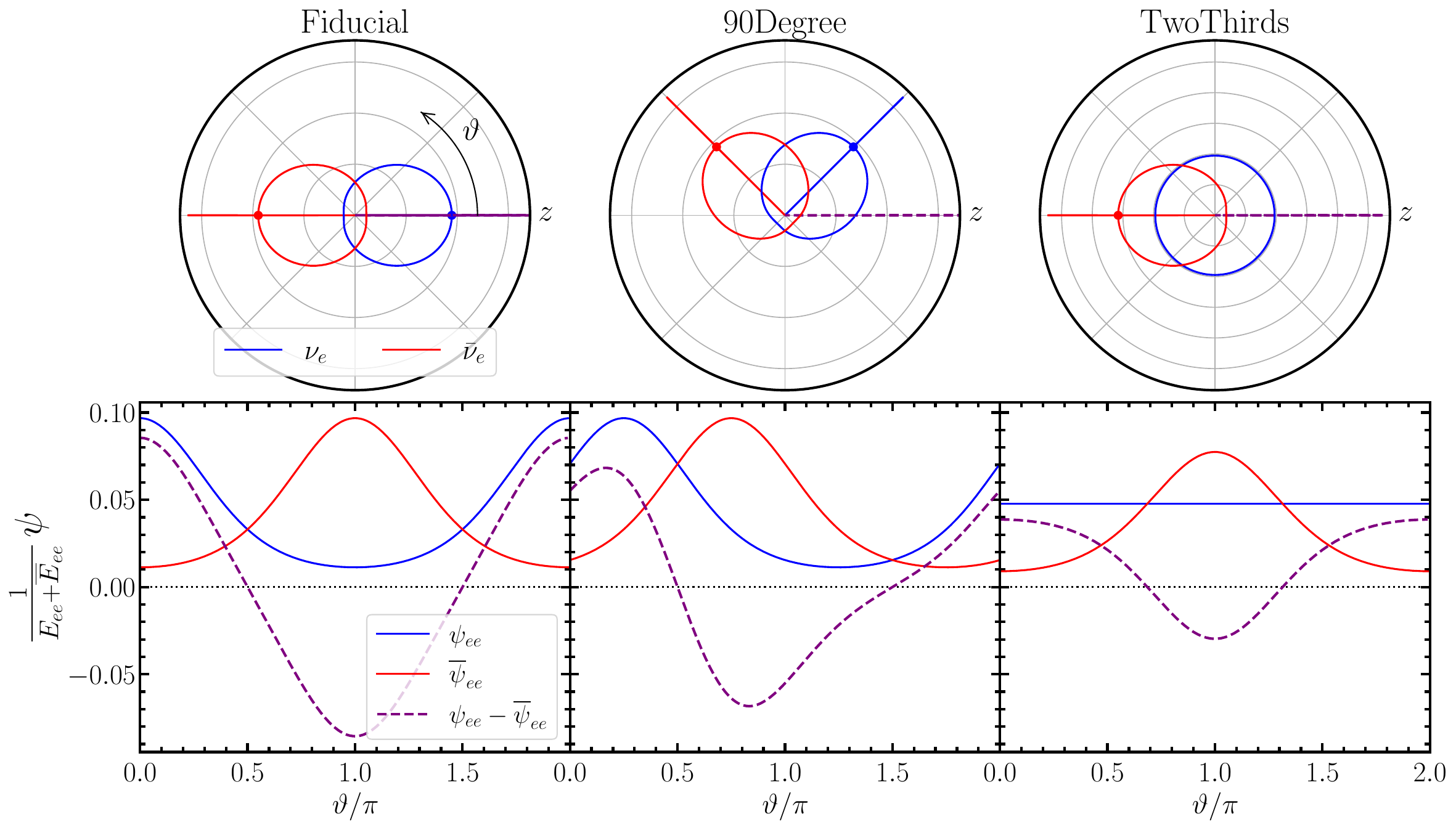}
    \caption{[Top] Polar representations of angular distributions for electron neutrino (blue) and electron anti-neutrinos (red) for the three tests at the beginning of the simulation.  Blue (red) vectors indicate the net flux direction.  Purple vector is the difference of blue and red vectors.  [Bottom] Curves as given by Eq.\ \eqref{eq:dn_dOmega} for $\nu_e$, $\overline{\nu}_e$ and the difference (purple) as a function of polar angle $\vartheta$. Angular distributions for $\nu_x$ and $\overline{\nu}_x$ are zero at the beginning of the simulation.  Lepton number crossings occur when purple line crosses $0$.
    }
    \label{fig:eln_3test}
\end{figure*}

The three 3D test problems we consider are named as Fiducial, 90Degree, and TwoThirds, all of which are described in detail in \citet{2021PhRvD.103h3013R, 2021PhRvD.104j3023R}. None of the three tests have analytic solutions\footnote{The Fiducial calculations of \cite{2021PhRvD.103h3013R} assume a slightly different angular distribution.}, so the comparison is based on how well the moment method of \flash can reproduce the PIC results.

\paragraph{Test parameters} Table \ref{tab:3test_parameters} gives the initial conditions for simulation parameters of the three tests.  The first three columns of Table \ref{tab:3test_parameters} give the initial values of the flavor-diagonal number density moment.  All three tests start with non-zero numbers of electron neutrinos and anti-neutrinos, and zero other-flavor neutrinos.  The fourth through sixth columns give the flux factor vectors.  Although these particular flux factor vectors need at most 2 dimensions to be fully described, we stress that the calculations are three dimensional and individual cells will generally develop flux moments where all three spatial components are non-zero.  The seventh column gives the side length of the domain, and the eighth column the number of cells.
Under the MEC, the initial angular distributions of all three tests in Table \ref{tab:3test_parameters} exhibit an ELN crossing and are therefore unstable to FFC.  

To visualize the geometry of these three tests, Fig.\ \ref{fig:eln_3test} shows the neutrino angular distributions [Eq.\ \eqref{eq:dn_dOmega}] for the electron neutrinos (blue) and anti-neutrinos (red).
The MEC distributions are 3D as emphasized above and as a result, we plot polar representations of 2D cross-sectional slices in the top row of Fig.\ \ref{fig:eln_3test}.  We measure the polar angle $\vartheta\in[0,2\pi]$ counter-clockwise from the $\widehat{z}$ axis.
We take the slices such that the maximum values of the distributions (that is the directions of the fluxes $\mathbf{F}_{ee}$ and $\overline{\mathbf{F}}_{ee}$) lie in the same plane. The polar plots in the upper panels show a more intuitive representation of the magnitude of the distribution in different directions, but the size and depth of the ELN crossings are more apparent in the standard plots on the lower panel. The vectors originating from the origin on the polar plots show the peak direction of the distributions. The difference of the blue and red vectors is shown in dashed purple — for instance, it is coincident with the blue vector on the Fiducial case, and the vector difference is shown vividly in the 90Degree test.  For all three tests, the coordinates are chosen so that the lepton number flux (i.e., the purple vector) lies along the $z$ axis. We then orient the polar plane so that this axis points in the rightwards direction, and indicate this direction as $\vartheta=0$ in the lower plots.  As in the polar plots, the blue and red curves give the electron neutrino and the electron anti-neutrino distributions.  Here, the purple curve gives the ELN distribution, properly normalized by the sum of the energy density moments.  As can clearly be seen in all three tests, the purple curves cross the horizontal axis implying a lepton number crossing.

\begin{figure*}[htb!]
    \centering
    \includegraphics[width=0.95\linewidth]{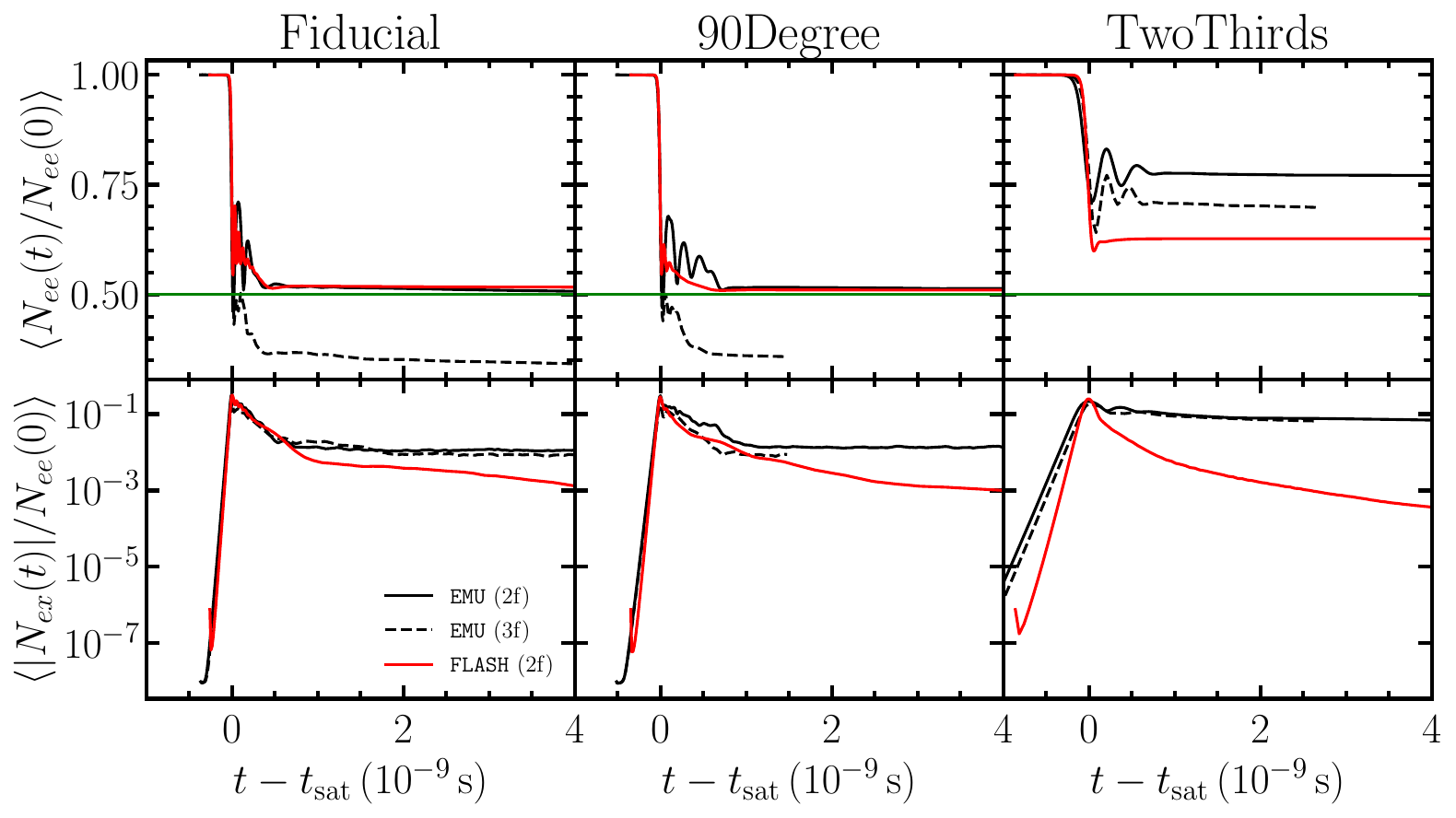}
    \caption{Density matrix elements versus time for the three tests. The horizontal axis is $t-t_\mathrm{sat}$ where $t_{\rm sat}$ differs between test case and method of calculation.  We use this definition for visualization purposes and stress the calculations are not simultaneous \revision{(see Fig.~\ref{fig:app_3test} for the same plots using directly the simulation time for the horizontal axis).} Two-flavor \flash quantities are plotted in red.  Two- (Three-) flavor \emu quantities are plotted in solid (dashed) black.  All quantities are averaged over the spatial domain, and in addition over particle number for \emu.  [Top] Plotted is the $ee$ component of the number density moment flavor-matrix $N$ (i.e., number density of $\nu_e$) scaled by $N_{ee}(t=0)$. [Bottom] Plotted is the magnitude of the off-diagonal component of $N$ scaled by $N_{ee}(0)$. For 3-flavor \emu calculations, we take the $e\mu$ component of $N$.
    }
    \label{fig:dm_tests}
\end{figure*}

\paragraph{Time evolution and FFI} We show the time-evolution of the domain-averaged values of $N_{ee}(t)/N_{ee}(0)$ (top) and $|N_{ex}(t)|/N_{ee}(0)$ (bottom) in Fig.\ \ref{fig:dm_tests} for \flash, 2-flavor \emu, and 3-flavor \emu simulations.
We emphasize that, given the different initial perturbations for \flash and \emu calculations (see Section~\ref{subsec:initial_boundary_conditions}), we do not expect identical time-evolution.  Subsequently, the saturation time, $t_{\rm sat}$, when the off-diagonal terms saturate (located at the peak of the $|N_{ex}|/N_{ee}(0)$ curves) depends on the initial conditions, as well as the kind of calculation.  To aid in visualization when comparing the growth, saturation, and decoherence phases between \flash and \emu, we define the horizontal axes in Fig.\ \ref{fig:dm_tests} as $t-t_\mathrm{sat}$ using a different $t_{\rm sat}$ for each calculation. We stress that none of the calculations are simultaneous with one another in simulation time -- the alignment at $t-t_{\rm sat}=0$ is a construct of the plot.
Finally, we plot a horizontal green line on the top panel to indicate the expected number of electron neutrinos in a \emph{2-flavor} calculation if the system were to completely mix flavor.

In all three tests, the \flash simulations exhibit fast flavor instability with a growth rate very similar to the true value. Considering the red (\flash) curves in Fig.\ \ref{fig:dm_tests}, there exists a period of exponential growth in $\langle|N_{ex}|\rangle$, evidencing one of the defining characteristics of the FFI.  $\langle|N_{ex}|\rangle$ continues to grow until the off-diagonal magnitude reaches the same order of magnitude as the initial electron-flavor number density moment.  When $|N_{ex}|\lesssim N_{ee}$ at saturation, there are rapid oscillations in the diagonal components, evidencing the other defining characteristic of FFI.  Saturation is a nearly instantaneous event with decoherence, i.e., decreasing $|N_{ex}|$, succeeding the rapid oscillations.  The decoherence continues as oscillations damp, with an end result of $\langle N_{ee}\rangle$ approaching an asymptote at a value less than the starting condition.
In summary, the results presented in Fig.\ \ref{fig:dm_tests} are quite remarkable in that even though instability criterion in FFI depends upon angular crossings of the ELN, the two-moment method accurately showcases the growth of the FFI without access to crossing information. Of course, crossings are implied by the MEC distribution used to generate the closure relation, but the MEC distribution is nowhere explicitly used in the code.

The growth rate in the \flash simulations is quantitatively very similar to that in the full \emu simulations, and even the final asymptotic neutrino distributions match well in certain cases.  The Fiducial and 90Degree tests show strong agreement between the two methods.  We see nearly identical growth rates for both tests, with \flash producing a slightly higher \imo.  Specifically for the Fiducial test case, $\imo=7.1\times10^{10}\,{\rm s}^{-1}$ in \flash, as compared to $\imo=6.3\times10^{10}\,{\rm s}^{-1}$ in \emu, for a difference of $\sim10\%$.  Results are similar for the 90Degree test, with $\imo=5.4\,(4.4)\times10^{10} \,{\rm s}^{-1}$ in \flash (\emu).  

Also, the asymptotic values for $\langle N_{ee}\rangle$ are nearly the same, with differences of $\sim1\%$ for both tests.  We nevertheless see differences between these two tests in the saturation and decoherence periods.  The oscillations in the top panels of Fig.\ \ref{fig:dm_tests} for \emu appear to have larger amplitudes and persist longer than those of \flash. Note that during the post-saturation decoherence, there appear to be two phases indicated by different slopes in $N_{ex}$. In the first, immediately after saturation, $N_{ex}$ decreases rapidly, but then numerical artifacts take over and decrease the decoherence rate (e.g., at around $t-t_\mathrm{sat}=0.75\,\mathrm{ns}$ in the Fiducial case). In the case of the \flash calculations, this is due to the numerical diffusion from finite grid spacing, and in the case of \emu this is due to the finite number of computational particles that achieve a state of random uncorrelated fluctuations, the amplitude of which scale very slowly as $N_p^{-1/2}$. Therefore the decoherence phase right after saturation is a robust physical prediction, but the late-time values of $N_{ex}$ show numerical artifacts.

The small amplitude oscillations for \flash are especially evident in the TwoThirds test case.  Here, we see a noticeable difference between \flash and \emu for the asymptotic values of $\langle N_{ee}\rangle$.  The growth rate is faster for \flash by $\sim40\%$ and the loss of coherence falls off faster.  There is a smaller amount of time when $|N_{ex}|\lesssim N_{ee}$, and thus less oscillations in the flavor-diagonal term.  The result is an asymptotic value which is $\gtrsim10\%$ of $N_{ee}(0)$.

For the TwoThirds test, we speculate that the reason the moment calculations do not asymptote at large times to the same value of $\langle N_{ee}\rangle$ as found by \emu is due to our imposition of the MEC.  Recall that for the \flash calculations, we use the quantum implementation of the MEC at every time step and substep of the evolution.
In contrast, the \emu calculations use Eq.\ \eqref{eq:dn_dOmega} only when generating the initial conditions, and the future evolution depends directly on the general distribution.
There is no guarantee that the neutrino distributions in \emu follow the MEC at any point except for initialization.

Although we have argued above that the use of the MEC in \flash necessarily restricts the shape the distributions may take during flavor evolution, there does exist the striking convergence between \flash and \emu of $\langle N_{ee}\rangle$ in the asymptotic limit for the Fiducial and 90Degree tests.
This is not a coincidence, but rather a result of the symmetry of both of these tests.
Initially, the system contains both $CP$ and rotational symmetries. The MEC is agnostic to $CP$ but does preserve the rotational invariance for constant flavor-traced flux factors.
As the energy density moment is equal for $E_{ee}$ and $\overline{E}_{ee}$, and the initial neutrino distributions are rotations of the anti-neutrino ones: an ELN crossing is inevitable.  The initial conditions and conservation of 3-momentum ensures that neutrinos and anti-neutrinos will never have the same flux factor vectors at any point in the test calculations.  As our system of equations is $CP$ symmetric (except for the initial conditions in the flux factors), we expect any flavor transformation for $E_{ee}$ to be accompanied by a commensurate change in $\overline{E}_{ee}$. Because 3-momentum is conserved, the flux factors are invariant and the ELN crossing persists to all times.  We have numerically verified that indeed $\overline{E}_{ee}$ mirrors the evolution of $E_{ee}$ and an ELN crossing exists in perpetuity.  In other words, the distributions shown in the top panels of Fig.\ \ref{fig:eln_3test} only scale in radial coordinate during their evolution.
However, the results in Fig.\ \ref{fig:dm_tests} clearly show a stable system post saturation.
For either the Fiducial or 90Degree system to become stable, an XLN crossing must develop, canceling the omnipresent ELN one \citep{2022PhRvL.129z1101N,2023PhRvD.107j3022Z,2023PhRvD.108f3003X}.
Furthermore, the $\nu_x$ and $\overline{\nu}_x$ distributions have the same vector flux factors and use the same flavor-traced flux factor, implying those distributions are identical to the ones in the top panels of Fig.\ \ref{fig:eln_3test} except for a difference in the radial coordinate.
In the presence of non-trivial ELN and XLN crossings, a zero net lepton number at all angles requires identical energy, flux, and pressure moments for the $xx$ components as compared to their $ee$ counterparts -- implying near flavor equilibration.  Even if the distributions do not follow Eq.\ \eqref{eq:dn_dOmega} and the MEC, the symmetry of the system guarantees that $\langle N_{ee}\rangle$ must converge to $50\%$ of the flavor trace [equivalent to $N_{ee}(0)/2$] in both the Fiducial and 90Degree tests.
This need not be the situation in the TwoThirds test case as the system neither exhibits $CP$ nor rotational symmetry.
Here, the MEC is not an accurate representation of the distributions at later times, and as a result, the \flash and \emu calculations show a stark divergence.
Discrepancies between moment and multi-angle methods were also seen in \cite{Myers}.

\begin{figure}[htb!]
    \centering
   \includegraphics[width=\columnwidth]{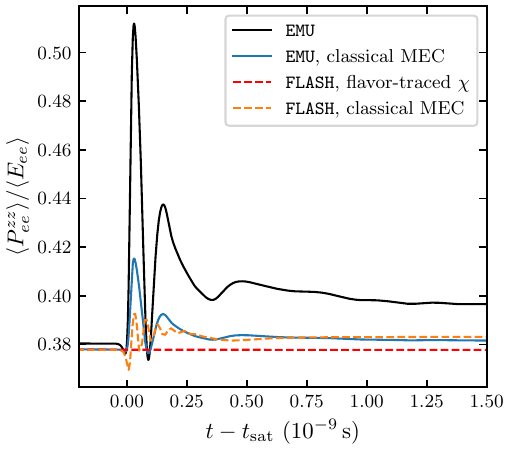}
    \caption{$zz$ component of the pressure tensor for electron-flavor neutrinos plotted against time for the Fiducial test case.  The pressure tensor component is normalized by the energy density moment.  The solid black curve gives the pressure tensor for the \emu simulation, while the dashed red curve gives the same for the \flash simulation.  Also included are diagnostic quantities for \flash (dashed orange) and \emu (solid blue) using a classical MEC along with the number and flux moments as given by the simulations.  All quantities are averaged over the simulation domain.  
    }
    \label{fig:pressure_fid}
\end{figure}

\paragraph{Pressure Moment}  As discussed above, the MEC need not be a true representation of the distribution even if we find flavor-convergence in the asymptotic limit. Figure \ref{fig:pressure_fid} gives a plot of the $zz$ pressure tensor component for the electron neutrinos in the Fiducial test case.  We pick the $zz$ component for $P_{ee}$ as $\widehat{z}$ is the direction of the net neutrino flux.  For the geometry of the Fiducial test case, the thin and thick components of the pressure tensor reduce simply to $P_{\rm thin}^{zz}/E=1$ and $P_{\rm thick}^{zz}/E=1/3$, implying that the interpolated value from Eq.\ \eqref{eq:Pij_closed} is $P^{zz}/E=\chi$.
For the purposes of analyzing our moment and PIC simulations, we plot the averaged values of $P_{ee}^{zz}/E_{ee}\sim\chi$ against the time as measured from the saturation peak. We choose this representation of our data as we do not expect qualitative differences for different cells.  In contrast, \citet{2023arXiv230814800N} plots the time-averaged values of the pressure against the radial coordinate when comparing multi-angle results to closure approximations in a global CCSN simulation, showing the transition from the optically thick to thin limit.
The solid black curve in Fig.\ \ref{fig:pressure_fid} corresponds to the baseline \emu calculation, i.e., the solid black curve in the upper-left plot of Fig.\ \ref{fig:dm_tests}.  To calculate $P^{zz}_{ee}$ for \emu, we use Eq.\ \eqref{eq:mom_2} to sum over the particles in a given cell and obtain the second angular moment of the distribution.  We subsequently average over the simulation domain and normalize by the energy density moment.  The dashed red curve gives the same quantity for the \flash simulation.  For \flash, we first calculate the domain average of the energy and flux density moments.  Along with the flavor-traced Eddington factor from Eq.\ \eqref{eq:f_FT}, we use Eq.\ \eqref{eq:Pij_closed} with the averaged $E_{ee}$ and $\mathbf{F}_{ee}$ to obtain $P^{zz}_{ee}$.  Finally, we normalize by $\langle E_{ee}\rangle$.  The constant value of the red dashed line shows that \flash is conserving both neutrino energy density (i.e., particle number) and neutrino flux density (i.e., 3-momentum).

For diagnostic purposes, we include two other pressure quantities in Fig.\ \ref{fig:pressure_fid}.  The dashed orange curve gives the pressure using the output \flash energy and flux moments along with the classical MEC prescription (i.e., an Eddington factor calculated using pure diagonal flux-factors without a trace over flavor). The solid blue curve gives the same but for output \emu quantities.  By comparing the blue and black curves, we see how much the distribution in the PIC calculation differs from the classical MEC.  Note that at times $t<t_{\rm sat}$, a finite number of particles causes the black curve to deviate from the blue one (we verified that increasing the number of particles reduces this discrepancy).  After saturation, the black curve exhibits a larger amplitude of oscillations as compared to the blue curve. For large values of $\langle P^{zz}_{ee}\rangle$, the actual PIC calculation is more forward peaked than the MEC approximation.  The opposite would be true for small values of $\langle P_{ee}^{zz}\rangle$, although it appears that the black and blue curves do not differ much at their minima.  This finding is consistent with \citet{2023arXiv230814800N} during periods of significant flavor transformation [see Figs.\ (8) and (9) of the aforementioned work], despite the differences in plotting axes. 
\citet{2023arXiv230814800N} show the closure relation cannot always describe the shape of the flavor-transformed distribution during rapid flavor oscillations as the Eddington factor falls outside of the classically allowed range.  Although $\langle P_{ee}^{zz}\rangle/\langle E_{ee}\rangle$ always falls within the classically allowed range for our Fiducial case, the difference between the blue and black lines is most acute at the maxima, and implies the MEC does not capture the multi-angle distribution at all times. Lastly, the black and blue curves oscillate nearly in phase with one another, indicating that the MEC contains the correct scaling of $P_{ee}^{zz}$ with $E_{ee}$ and $\mathbf{F}_{ee}$ but not the correct sensitivity.

We notice another difference in sensitivity when comparing the solid blue curve to the dashed orange curve of \flash.  The classical MEC calculation using \flash data shows a smaller amplitude of oscillation, along with a larger frequency.  We attribute the smaller amplitudes to the fact that the MEC underestimates the degree of forward-peaking of the distribution.  The larger frequency correlates with the smaller time scales exhibited by \flash, and observed in all three test cases.  Notice that the dashed orange and solid blue curves do asymptote to similar values during the decoherence period, implying that the zeroth and first moments have similar values between the two methods of calculation.  Finally, we note that our choice of utilizing the flavor-traced Eddington factor (dashed red curve in Fig.\ \ref{fig:pressure_fid}) over the classical MEC in the \flash simulation results in a value of $\langle P^{zz}_{ee}\rangle$ differing by $\sim1\%$ of $\langle E_{ee}\rangle$.  As we operate in the optically thick limit at all times for this test case, we do not foresee that adopting the classical MEC prescription for calculating $\chi$ would alter the results in Fig.\ \ref{fig:dm_tests} by more than a few percent.

\begin{figure*}[htb!]
    \centering
    \includegraphics[width=\linewidth]{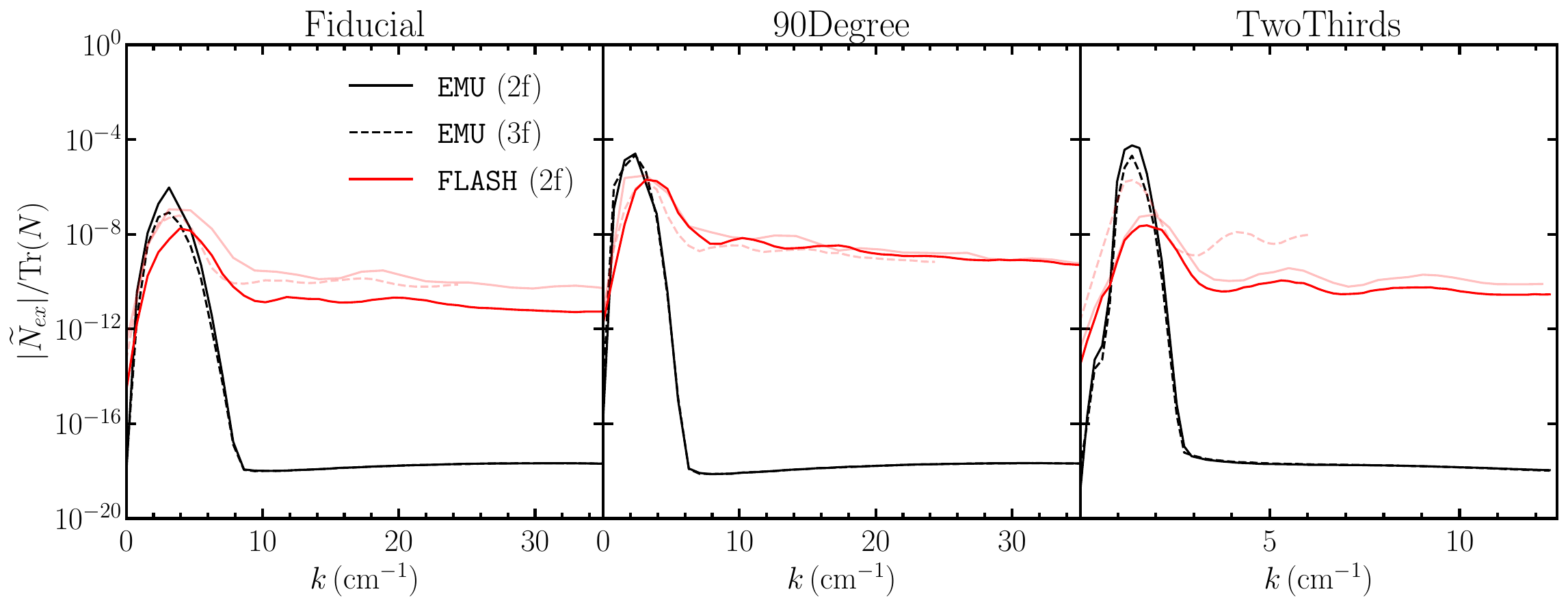}
    \caption{Magnitude of the discrete Fourier transform of $N_{ex}$ for all three tests plotted against wavenumber.  The DFTs are calculated at a time prior to saturation.  The DFTs for the Fiducial and 90degree test cases are taken $0.1\,{\rm ns}$ before saturation, and TwoThirds $0.35\,{\rm ns}$ before saturation.
    2-flavor (3-flavor) \emu simulations correspond to the black solid (dashed) curve; \flash simulations to the red solid curve.  The two light-red curves correspond to different resolution tests for \flash. The light-red solid curve has half the box-side length and half the number of grid points compared to columns 7 and 8 in Table \ref{tab:3test_parameters}.  The light-red dashed curve has the same box-side length and half the number of grid points.
    }
    \label{fig:FFTtests}
\end{figure*}

\paragraph{Fourier space analysis} We have discussed averages of the number density and pressure moments when presenting Figs.\ \ref{fig:dm_tests} and \ref{fig:pressure_fid}.  In Fig.\ \ref{fig:FFTtests} we show information on the structure in the simulation domain by using Discrete Fourier Transforms (DFTs).  The horizontal axes give the wavenumber $k$, and the vertical axis the magnitude of the DFT of the complex flavor-off-diagonal number density moment $N_{ex}$, normalized by the flavor trace.  We will refer to wavenumber values as ``modes''.  The solid red, solid black, and dashed black lines all correspond to the same simulations as Fig.\ \ref{fig:dm_tests}.  In lighter shades of red we have plotted DFTs from two additional \flash calculations of the same test cases. The light-red solid curve is from a simulation with the same number of grid points per cm but with a box-side-length of half the original simulation compared to the values in columns 7 and 8 of Table \ref{tab:3test_parameters}; the light-red dashed curve is from a calculation with the same domain size, but half the number of grid points per cm and a smaller maximum value of $k$.

The three DFTs in each panel of Fig.\ \ref{fig:FFTtests} are all from a time before saturation during the growth period: $\sim0.1\,{\rm ns}$ before saturation for the Fiducial and 90Degree cases; $\sim0.35 \,{\rm ns}$ before saturation for the TwoThirds case. While similar, the times of the snapshots used in the DFTs are not exactly equal between different calculations so the values of $\widetilde{N}_{ex}$ cannot be compared across either the simulations or the resolution tests. For this reason, comparisons should be restricted to within an individual calculation, i.e., the relative heights of peaks.  

The DFTs show the scales, via wavenumber $k$, where there exists a sinusoidal pattern in the flavor off-diagonal number density moment.  This superposition of sinusoids need not have growing amplitudes for each mode.  A priori, only one mode is necessary to explain the growth phase in Fig.\ \ref{fig:dm_tests}.  However, during the growth phase, all modes in Fig.\ \ref{fig:FFTtests} do indeed grow in power until saturation, implying there are many unstable modes in the system.

All three tests show a discernible peak in the dark red curves of Fig.\ \ref{fig:FFTtests}.  Soon after the simulations begin the DFT exhibits a peak with an associated wave number as evidenced in Fig.\ \ref{fig:FFTtests}. The peak remains at that location in $k$, although with growing height, until saturation.
The DFTs for the resolution tests show similar behavior in the peak position and growth phase, indicating that the dark red curve for the simulation is indeed spatially resolved.
We call the wavenumber at this peak the fastest growing mode \kmax.  The growth rate in Fig.\ \ref{fig:dm_tests} is linked to the fastest growing mode via a dispersion relation, with details provided in \citet{Froustey:2023skf}.

The \flash and \emu calculations both have discernible peaks with similar fastest growing modes.  The wave number of the fastest growing modes for \flash are slightly larger, reflecting a smaller scale.  For example, $\kmax=3.9\,(3.1)\,{\rm cm}^{-1}$ for \flash (\emu) in the Fiducial test case, and $\kmax=3.1\,(2.4) \, \mathrm{cm^{-1}}$ in the 90Degree case.  Also, it appears that the noise floor of the DFT is larger for \flash, or equivalently, there exists relatively less power in the fastest growing mode.  Lastly, there are a few harmonics visible in \flash but not present in \emu.  This is true for all three test cases, and more pronounced for the TwoThirds case.  These harmonics, however, only crest slightly above the noise floor.

\begin{table*}
    \centering
    \hspace{-1.5cm}
    \begin{tabular}{c|ccccc}
        \multirow{2}{*}{Name} &  \multirow{2}{*}{$\langle|N_{ex}|/N_{ee}(0)\rangle|_{t=t_{\rm sat}}$} & \multirow{2}{*}{$\langle|N_{ex}(t_{\rm dec})|\rangle/\langle|N_{ex}(t_{\rm sat})|\rangle$} & \multirow{2}{*}{$\langle N_{ee}/N_{ee}(0)\rangle|_{t\rightarrow\infty}$} & \imo & \kmax \\
         &&&& $(10^{10}\,\mathrm{s}^{-1})$ & $(\mathrm{cm}^{-1})$ \\\hline
        Fiducial & & & & & \\
        \flash     & $0.314$ & $0.292$ & $0.517$ &
        $7.1$ &
        $3.9(4)$\\
        \emu (2f)  & $0.333$ & $0.360$ & $0.506$ &
        $6.3$ &
        $3.1(4)$\\ \hline
        90Degree & & & & & \\
        \flash     & $0.281$ & $0.191$ & $0.510$ &
        $5.4$ &
        $3.1(4)$\\
        \emu (2f)  & $0.303$ & $0.333$ & $0.516$ &
        $4.4$ &
        $2.4(4)$\\ \hline
        TwoThirds & & & & & \\
        \flash     & $0.248$ & $0.216$ & $0.627$ &
        $2.0$ &
        $1.8(1)$\\
        \emu (2f)  & $0.214$ & $0.579$ & $0.771$ &
        $1.2$ &
        $1.4(1)$\\ \hline
    \end{tabular}
    \caption{Numerical results for \flash and \emu (2f) calculations for the three 3D test simulations. First column gives the ratio $\langle \lvert N_{ex} \rvert/N_{ee}(0)\rangle$ when $t=t_{\rm sat}$.
    Second column gives the ratio of off-diagonal magnitudes at two different times: $t_{\rm sat}$ and $t_{\rm dec}=t_{\rm sat}+0.2\,{\rm ns}$. 
    Third column gives the asymptotic ratio $\langle N_{ee}/N_{ee}(0)\rangle$ post-saturation. Fourth column gives the growth rate \imo when the system is unstable in units of $10^{10}\,{\rm s}^{-1}$.  Last column gives the fastest growing mode in the domain \kmax in units of ${\rm cm}^{-1}$, with an associated uncertainty in parentheses.  
    The values of the first two columns are much more variable with the initial random perturbations than the last three.
    }
    \label{tab:3test_results}
\end{table*}

In summary, Table \ref{tab:3test_results} gives numerical results of FFC to compare between \flash and the 2-flavor \emu calculations for all three tests.
The values in columns one through four are deduced from Fig.\ \ref{fig:dm_tests} and are the following, respectively: the maximum value of $\langle|N_{ex}|/N_{ee}(0)\rangle$ in the bottom panels; the ratio of $\langle |N_{ex}|\rangle$ at the saturation time and a time $t_{\rm dec}=t_{\rm sat}+0.2\,{\rm ns}$ during the decoherence phase; the asymptote of $\langle N_{ee}/N_{ee}(0)\rangle$ in the top panels; and the slope of the line (in semi-log space) in the bottom panels.  The fifth column gives the value of $k$ at the peak of the DFT in Fig.\ \ref{fig:FFTtests}. We give an uncertainty in parentheses for \kmax due to the finite box size, namely, $\delta k=\pm\pi/L$.
All tests show the \flash calculations have larger values of \imo and \kmax compared to \emu.
In addition, the rate of decline of $|N_{ex}|$ is larger for the \flash calculations in all three tests.
However, even with the different growth and loss of coherence rates, oscillations occur while the average value of $|N_{ex}|$ exhibits quite similar values for each test.
This value, in the first column, is similar within a given test but not uniform across all three tests. Moreover, it varies with the random initial perturbations and should not be taken as a robust prediction for each calculation, contrary to the growth rate, instability lengthscale, and amount of flavor transformation.

\subsection{Neutron Star Merger}
\label{sec:results_nsm}

Our next set of simulations use initial conditions extracted from the three-dimensional neutron star merger simulation of \cite{Foucart:2016rxm}.
This simulation is general relativistic, but simulating neutrino oscillations in curved spacetime is 
beyond the scope of this work. Appendix \ref{sec:ortho} describes our procedure on transforming the 
distributions defined in a general spacetime metric to distributions defined in an orthonormal tetrad comoving with the fluid. In this frame, the construction of a flavor transformation simulation is more intuitive, since we can treat the spacetime as locally flat.

\begin{figure}[htb!]
    \centering
    \includegraphics[width=\linewidth]{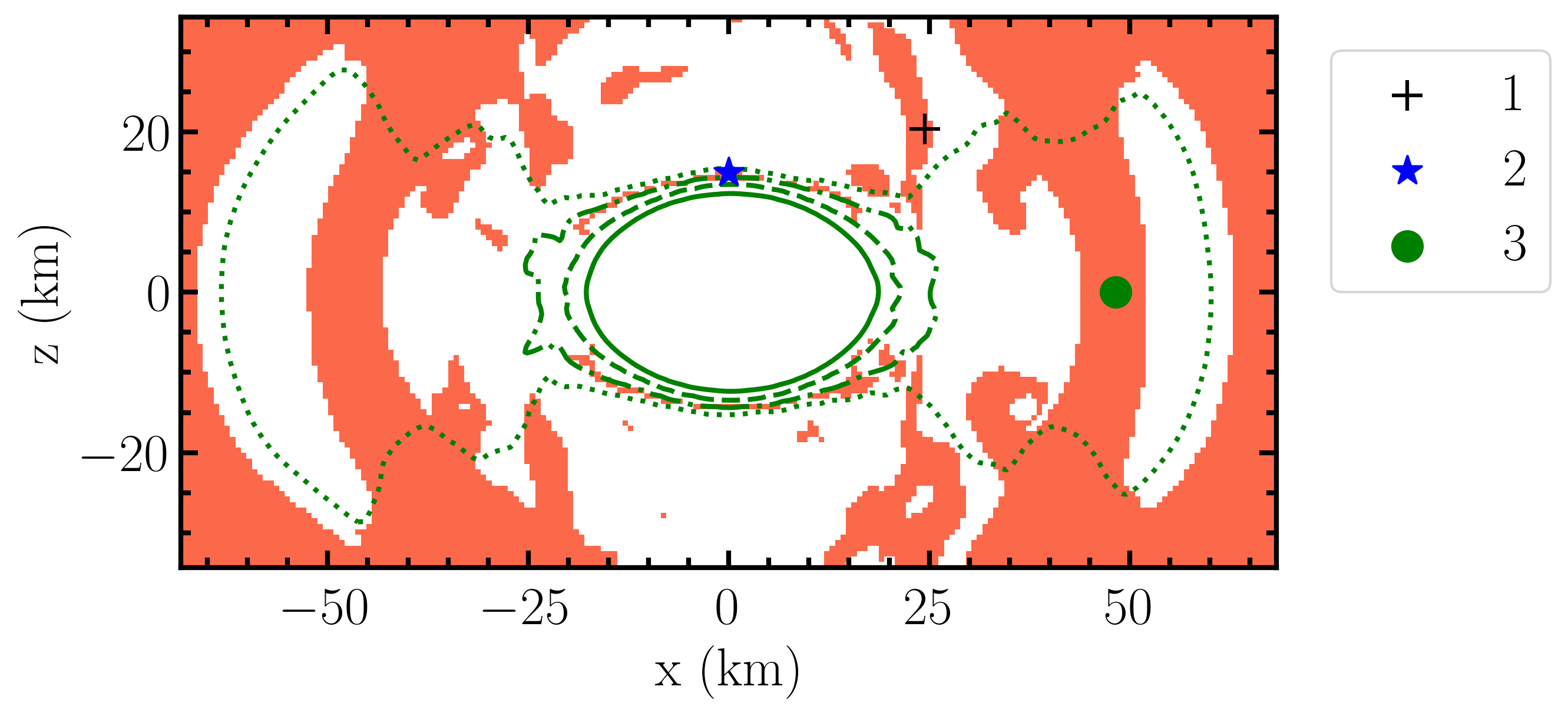}
    \caption{NSM crossing information from \citet{Foucart:2016rxm}. The snapshot is taken $5\,{\rm ms}$ post merger. Red pixels indicate locations where an ELN crossing exists. The three symbols (black cross, blue star, green circle) indicate the locations for the three flavor-transformation simulations we consider in Sec.\ \ref{sec:results_nsm}. From outside to inside, the green contours indicate matter densities of \{$10^{11},10^{12},10^{13},10^{14}$\} $\mathrm{g.cm}^{-3}$.
    }
    \label{fig:foucart}
\end{figure}

We analyze and simulate neutrino distributions at a selection of points in the polar slice of a snapshot at $5\,{\rm ms}$ post merger shown in Fig.\ \ref{fig:foucart}.
Green contours give matter densities of \{$10^{11},10^{12},10^{13},10^{14}$\} $\mathrm{g\,cm}^{-3}$ and the inner contours show the position of the central compact object at the center of the domain.  The red pixels indicate where an ELN crossing exists according to 
Eq.\ \eqref{eq:eln_cross}.  White pixels indicate that no such ELN crossing exists, although these regions 
are still subject to flavor transformation via other processes (e.g., the matter-neutrino resonance) or advection of flavor-transformed distributions into those regions of space. We select three points to simulate in \flash and \emu, indicated by the black cross, blue star, and green circle, to model regions above the accretion disk, within the disk, and just outside of the compact object, respectively.
The black cross is the same point detailed in \citet{Grohs:2022fyq}.

\renewcommand{\arraystretch}{1.2}

\begin{table*}
    \centering
    \hspace{-2.5cm}
    \begin{tabular}{c|cccccccc}
        \multirow{2}{*}{Name} & $\mathcal{N}_{ee}$ & $\overline{\mathcal{N}}_{ee}$ & $\Sigma \mathcal{N}_{(x)}$ & $\mathbf{f}_{ee}$ & $\overline{\mathbf{f}}_{ee}$ & $\mathbf{f}_{(x)}$ & $L$ & $N_{gp}$\\
        & $(10^{32}\mathrm{cm}^{-3})$ & $(10^{32}\mathrm{cm}^{-3})$ & $(10^{32}\mathrm{cm}^{-3})$ & &&& (cm)\\\hline
        NSM 1 & 14.22 & 19.15 & 19.65
        & $\begin{pmatrix}\phantom{-}0.0974\\ \phantom{-}0.0421\\ -0.1343\end{pmatrix}$
        & $\begin{pmatrix}\phantom{-}0.0723\\ \phantom{-}0.0313\\ -0.3446\end{pmatrix}$
        & $\begin{pmatrix}-0.0216\\ \phantom{-}0.0743\\ -0.5354\end{pmatrix}$
        & 7.87 & $128^3$\\
        NSM 2 & 23.29 & 28.53 & 60.11
        & $\begin{pmatrix}\phantom{-}0.0086\\ -0.0174\\ -0.1635\end{pmatrix}$
        & $\begin{pmatrix}\phantom{-}0.0070\\ -0.0142\\ -0.2338\end{pmatrix}$
        & $\begin{pmatrix}-0.0476\\ -0.0231\\ -0.2679\end{pmatrix}$
        & 8.27 & $256^3$\\
        NSM 3 & 28.80 & 37.42 & 19.32
        & $\begin{pmatrix}\phantom{-}0.0004\\ -0.0033\\ \phantom{-}0.0044\end{pmatrix}$
        & $\begin{pmatrix}\phantom{-}0.0003\\ -0.0025\\ -0.1306\end{pmatrix}$
        & $\begin{pmatrix}-0.0008\\ -0.0051\\ -0.1292\end{pmatrix}$
        & 5.80 & $512^3$ \\
    \end{tabular}
    \caption{List of baseline simulation parameters for the \flash NSM simulations. Column labels are the same as Table \ref{tab:3test_parameters}. Note that all corresponding \emu simulations were run with the same parameters, but using $N_{gp}=128^3$ grid cells due to the longer wavelength of the fastest growing mode.
    }
    \label{tab:NSM_parameters}
\end{table*}

\renewcommand{\arraystretch}{1.}

\begin{figure*}[htb!]
    \centering
    \includegraphics[width=\linewidth]{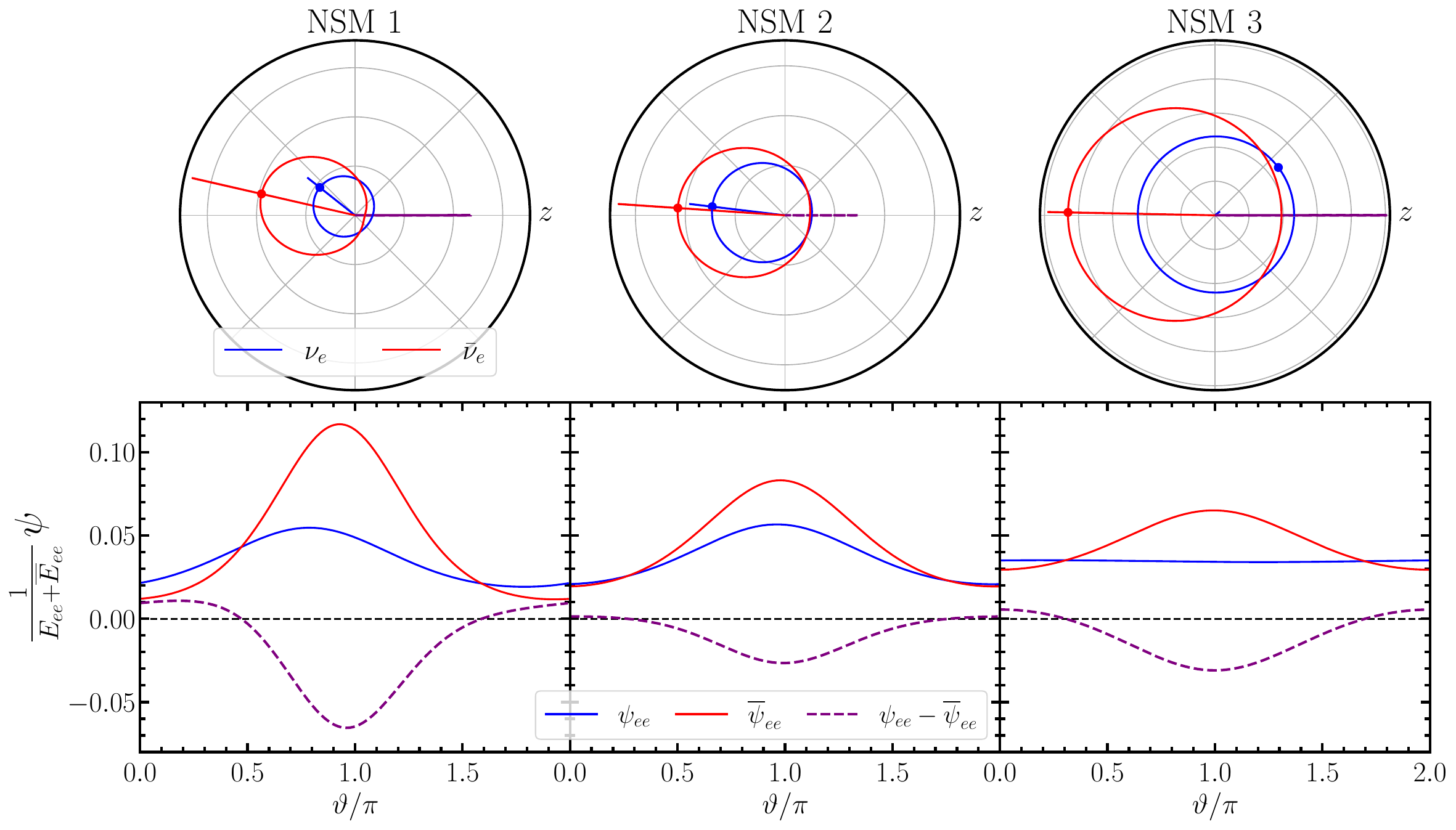}
    \caption{Polar and cartesian representations of the initial $\nu_e$ and $\overline{\nu}_e$ distributions for the NSM points. Plotting conventions are the same as Fig.\ \ref{fig:eln_3test}. Note that the $z$ axis, corresponding to $\vartheta = 0$, is a local coordinate chosen differently at each NSM point to be coincident with the lepton number flux direction.
    }
    \label{fig:eln_3NSM}
\end{figure*}

Table \ref{tab:NSM_parameters} and Fig.\ \ref{fig:eln_3NSM} are the NSM analogs to Table \ref{tab:3test_parameters} and Fig.\ \ref{fig:eln_3test} of Sec.\ \ref{sec:results_tests}, and use the same notation and plotting conventions.  Note that the orthogonalization procedure in App.\ \ref{sec:ortho} is location-dependent and as a result the directions of the fluences in Table \ref{tab:NSM_parameters} cannot be compared to one another between points.  In other words, for this particular study our flavor transformation simulations are restricted to the local area of each point and do not affect one another through advection. However, the neutrinos for points 1 and 2 are generally moving upward, so the leftward direction in the polar plots in Fig.\ \ref{fig:eln_3NSM} roughly correspond to the $\hat{\mathrm{z}}$ direction in Fig.\ \ref{fig:foucart}. The leftward direction in the polar plot for point 3 in Fig.\ \ref{fig:eln_3NSM} roughly corresponds to the $\hat{\mathrm{x}}$ direction in Fig.\ \ref{fig:foucart}. Note that while there are healthy ELN crossings in points 1 and 3, the crossings in point 2 are quite tenuous, as would be expected given the very thin band of instability just above the compact object in Fig.\ \ref{fig:foucart}.

\begin{figure*}[htb!]
    \centering
    \includegraphics[width=0.92\linewidth]{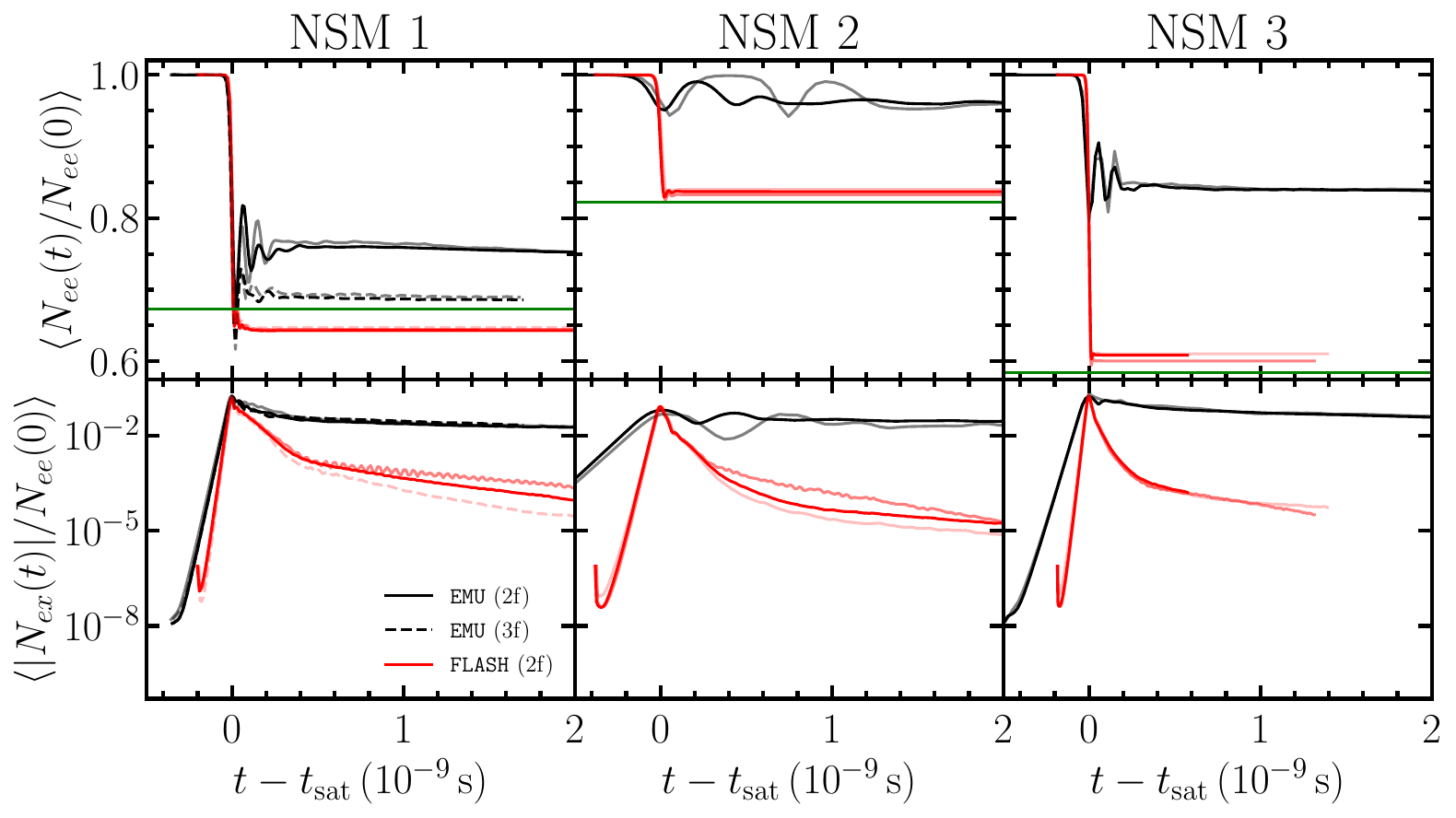}
    \caption{Domain averaged components for the number density moment plotted against time measured from the point of saturation. $t_{\rm sat}$ differs between calculations in the same manner as Fig.~\ref{fig:dm_tests} \revision{(see Fig.~\ref{fig:app_3NSM} for the same plots using directly the simulation time for the horizontal axis)}.  Panels and curve conventions are similar to Fig.\ \ref{fig:dm_tests} and simulation computational parameters are given in the last two columns of Table \ref{tab:NSM_parameters}.
    Gray and light red lines give results from resolution tests and are dependent on the NSM point.
    For NSM 1, the solid (dashed) gray lines are for \emu 2-flavor (3-flavor) simulations with half the side-length and half the number of grid points per side, for an identical grid spacing.  The solid medium-red line also is for a \flash simulation with half the side-length and half the number of grid points.  In addition, the dashed light-red line is for a simulation with half the side-length but the same number of grid points, for half the grid spacing.
    For the 3-flavor \emu calculation in NSM 1, $ex=e\mu$.
    NSM 2 and 3 follow identical conventions for resolution testing compared to one another.  Gray lines are for 2-flavor \emu simulations with half the domain size and half the number of grid points per side.  Medium-red \flash simulations are also half the domain size and half the number of grid points per side.  The light-red \flash calculation is for the same domain side-length, but only half the number of grid points resulting in twice the grid spacing of the baseline simulation. 
    }
    \label{fig:P_NSM}
\end{figure*}

\paragraph{Time evolution and FFI} Figure \ref{fig:P_NSM} shows the time-evolution of the neutrino number density moment for all three NSM points. Contrary to the previous test cases, the conditions in the NSM dictate non-zero initial distributions of heavy lepton neutrinos.
This results in different 2-flavor complete mixing
lines for each simulation, shown in green in Fig.\ \ref{fig:P_NSM}.  The lighter-opacity lines are for different resolution tests (see descriptions in caption).  
For illustrative purposes, we also include a three flavor EMU simulation for the first NSM point and plot the $e\mu$ component in the bottom panel.

In all three NSM points, we see growth, saturation, and decoherence phases as we did in Sec.\ \ref{sec:results_tests} and Fig.\ \ref{fig:dm_tests}.
Growth of $|N_{ex}|$ begins soon after the start of the simulation and proceeds until $\langle|N_{ex}|\rangle\sim0.1\langle N_{ee}\rangle$ in both \flash and \emu calculations.  Rapid oscillations develop and effect a decrease in $\langle N_{ee}\rangle$ towards complete flavor-mixing.  Notice that for the first NSM point treated by \flash, the conditions are such that $\langle N_{ee}\rangle$ falls below the complete-mixing green line and asymptotes to a value less than $50\%$ of the flavor trace.  This is the case for all three resolution tests, including the light-red dashed curve with half the gird spacing compared to the baseline simulation. The \emu results also briefly dip below the 50\% line, and it seems that the more rapid decoherence in the moment method halts the flavor transformation before it can oscillate back up. In all other calculations (\flash and \emu) $\langle N_{ee}\rangle$ remains above the green line at all times.  Decoherence enters after saturation in much the same manner as the three test calculations in Sec.\ \ref{sec:results_tests}.  In the first and second NSM points, the baseline and resolution tests for \flash begin to lose convergence in the decoherence phase. The divergence occurs well after saturation and at a point where $\langle N_{ee}\rangle$ has reached a steady-state value.  The \flash baseline and resolution tests for the third NSM point maintain convergence longer -- a result of this set of calculations having smaller grid spacings compared to the other two points.

We identify some general trends in the \flash and \emu results. \flash generally shows
faster growth rates, faster decoherence fall-offs, and less oscillations in the $N_{ee}$ moment, similar to the test cases in the previous section.  
The discrepancies are particularly apparent for the second NSM point. This point is unique in that the distribution described by the MEC is only marginally unstable. This type of condition is expected to lead to slower growth rates, less total flavor transformation, and more dependence on details in the small angular region between the ELN crossing points (e.g., \citealt{2021PhRvD.103h3013R, Bhattacharyya:2022eed}).
Specifically for \flash (\emu), $\imo=5.2\,(1.1)\times10^{10}\,{\rm s}^{-1}$ for this point. 
The bottom panels of Fig.\ \ref{fig:eln_3NSM} shows that ELN crossings are initially present for all three points, but the crossing is most shallow for the second point.
The two moment method plus MEC we are employing in
\flash is not able to capture the FFI behavior as well for this scenario as it is for the more pronounced ELN of the first point, and seems to behave as if there was a more significant instability than present in the detailed angular distribution, although we again emphasize that \flash still demonstrates a characteristic evolution pattern for a fast-flavor-unstable distribution in general.

\begin{figure*}[htb!]
    \centering
    \includegraphics[width=\linewidth]{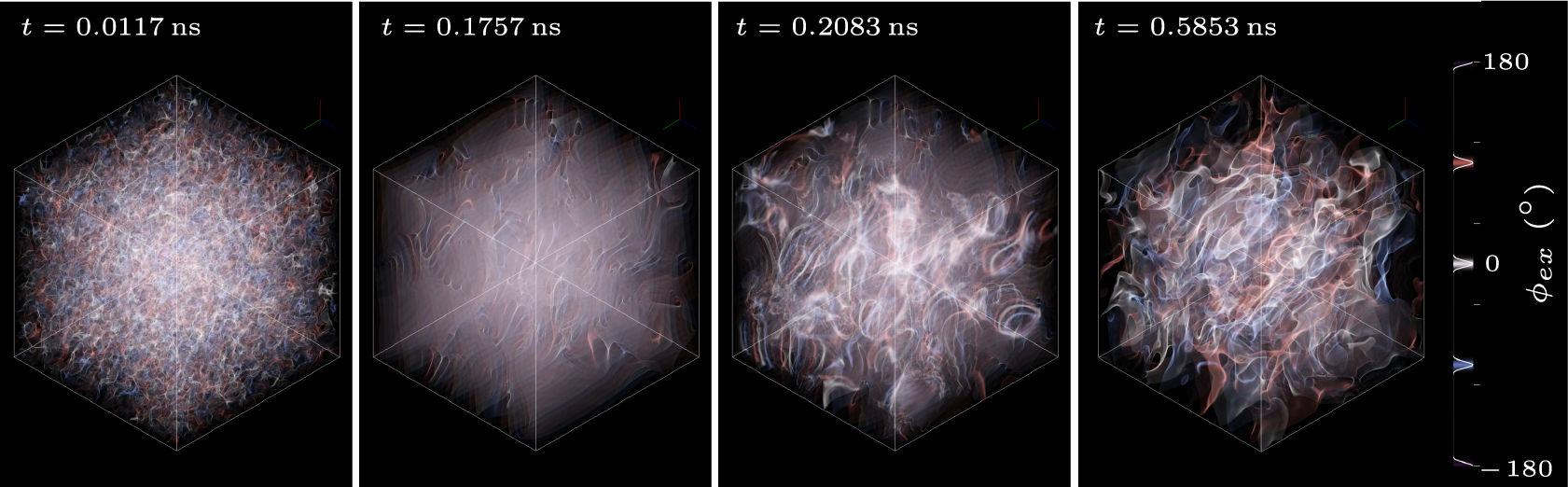}
    \caption{Volume rendering of contours of the phase of $N_{ex}$, $\phi_{ex}$, for NSM 1 (black cross in Fig.\ \ref{fig:foucart}). Blue, white, and red contours correspond to $\phi_{ex}=-\pi/2,0,\pi/2$, respectively.  The four panels are at four different times and roughly correspond to different phases of FFC. From left to right: $t=0.01\,{\rm ns}$ (initial conditions); $t=0.18\,{\rm ns}$ (growth phase); $t=0.21\,{\rm ns}$ (saturation point); and $t=0.59\,{\rm ns}$ (decoherence phase).
    }
    \label{fig:volume_rendering}
\end{figure*}

We show in Fig.~\ref{fig:volume_rendering} 3D-volume-renderings of the \flash simulation for the first NSM point at four different simulation times: $t=0.01\,{\rm ns}$; $t=0.18\,{\rm ns}$; $t=0.21\,{\rm ns}$; and $t=0.59\,{\rm ns}$.
The contours in each panel are for the phase, $\phi_{ex}$, of the complex number $N_{ex}$.
The spatial structure in $\phi_{ex}$ reflects the phase of the growing mode, and so reflects the three-dimensional structure of the peak of the DFT during the linear phase and a combination of the persisting mode structure and random decoherence after the instability saturates.
We plot three contours for the phase: $\phi_{ex}=-\pi/2$ (blue); $\phi_{ex}=0$ (white); and $\phi_{ex}=\pi/2$ (red).  The first panel shows a time close to the start of the simulation, where little flavor-transformation has occurred and the contours are close to the initial conditions of the random distributions with no structure [Eq.\ \eqref{eq:od_seed}].  We can see some structure in the second panel during the growth phase, where the distance between planar structures reflects the wavelength of the fastest growing mode and the planes are roughly perpendicular to the direction of the net ELN flux. The phases are distorted when the evolution becomes nonlinear as the instability saturates in the third panel.  The last panel is during the decoherence phase when the flavor field is no longer unstable, yet there still exists structure in $\phi_{ex}$.  The pattern seen here in Fig.\ \ref{fig:volume_rendering} is qualitatively similar to that seen in Fig.\ (2) of \citet{2021PhRvD.104j3023R}, albeit for a different simulation that employed the PIC method.  Nevertheless, the similarity in the growth of structure of $\phi_{ex}$ again shows that the moment method reproduces many features of the FFI on large and small scales.

\begin{figure}[htb!]
    \centering
   \includegraphics[width=\columnwidth]{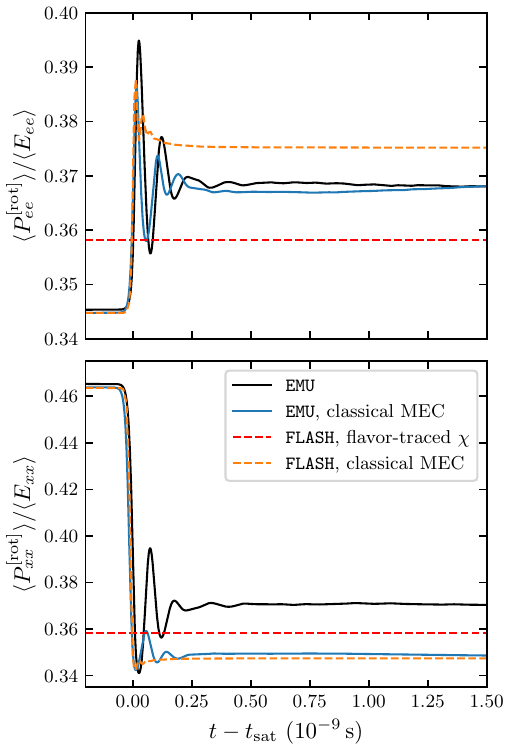}
    \caption{Component parallel to flux of pressure tensor plotted against time for the NSM 1 point.  Line and axes conventions are the same as Fig.\ \ref{fig:pressure_fid}.  Top panel gives electron flavor pressure tensor, while bottom panel gives heavy-lepton $x$ flavor.  The rotation of the pressure moment is different for each flavor and at each time step.
    }
    \label{fig:pressure_nsm1}
\end{figure}

\paragraph{Pressure Moment} In analogy to the Fiducial test case in Sec.~\ref{sec:results_tests}, we compare components of the pressure tensor between \flash and \emu calculations for the NSM 1 point in Fig.~\ref{fig:pressure_nsm1}. For the Fiducial test case, we chose the pressure component along the symmetry axis $\widehat{z}$ when drawing Fig.\ \ref{fig:pressure_fid}.  No such symmetry axis exists for the NSM 1 point, so we instead rotate into a primed reference frame where a principal axis is aligned with the flux vector for a given density matrix component.  In other words, if we define a basis $(\widehat{x}^{\prime},\widehat{y}^{\prime},\widehat{z}^{\prime})$ such that $\widehat{z}^{\prime}$ is the unit vector in the $\mathbf{F}_{aa}$ direction, we compute $P_{aa}^{[\mathrm{rot}]} \equiv P_{aa}^{z^{\prime} z^{\prime}}$, which we obtain after an appropriate spatial rotation of the pressure tensor. The rotation is different for each flavor, i.e., $a=e,x$, and computed at each time step.  The solid black curve represents this quantity, averaged over the simulation domain and normalized by $\langle E_{aa} \rangle$, for electron and heavy lepton flavor neutrinos. If the pressure moment is obtained from the closure relation~\eqref{eq:Pij_closed}, then by construction $P_{aa}^{[\mathrm{rot}]}/E_{aa} = \chi_{aa}$. This is indeed the case for the dashed red curve in the top and bottom panels of Fig.\ \ref{fig:pressure_nsm1}, which is equal to the flavor-traced Eddington factor used in \flash. For diagnostic purposes, we also represent the pressure moment computed using the classical MEC prescription, i.e., the non flavor-traced Eddington factor obtained from the first two angular moments in \emu (solid blue curve) and \flash (dashed orange curve).

Similar to the Fiducial test case and Fig.\ \ref{fig:pressure_fid}, the MEC is able to capture some of the features of the underlying distribution.  We note that the solid blue curve tracks the black curve quite closely and asymptotes to nearly identical values for the electron flavor pressure (similar to Fig.~\ref{fig:pressure_fid}, the initial discrepancy between the black and blue curves is due to the finite number of particles).  Similar to the Fiducial test case, the black curve tends to have more extreme maxima and minima, implying that the MEC underestimates the degree of forward-peaking \citep{2023arXiv230814800N}.  For \flash, the dashed orange curve follows the solid blue curve for roughly half a period during the onset of rapid flavor oscillations immediately after saturation.  These oscillations terminate prematurely for the \flash simulation and continue for \emu, implying final asymptotic values for $\langle P^{\rm [rot]}_{ee}\rangle$ which differ by few percent.  However, we observe that for the electron flavor component, there is little variation in $\langle P^{\rm [rot]}\rangle$ over the simulation time, and we remain close to the optically thick limit for its entirety.  The differences in the solid black, solid blue, and dashed orange lines are only a few percent compared to the flavor-traced Eddington factor displayed in the dashed red curve.

In the bottom panel of Fig.\ \ref{fig:pressure_nsm1}, we again see good agreement between the black, blue, and orange curves prior to saturation.  $\langle P^{\rm [rot]}_{xx}\rangle$ encompasses a larger range than $\langle P^{\rm [rot]}_{ee}\rangle$.  Unlike the electron flavor, here the blue and orange curves show the best agreement for the $x$ flavor, indicating that the \flash simulation captures the first two moments of $\varrho_{xx}$ in accordance with \emu over a large timespan.  Furthermore, regardless of the flavor, the orange and blue curves agree closely with the black curve in the growth phase.  Although we use a flavor-traced Eddington factor in \flash, the first two moments remain accurate in that linear phase.

\begin{figure*}[htb!]
    \centering
    \includegraphics[width=\linewidth]{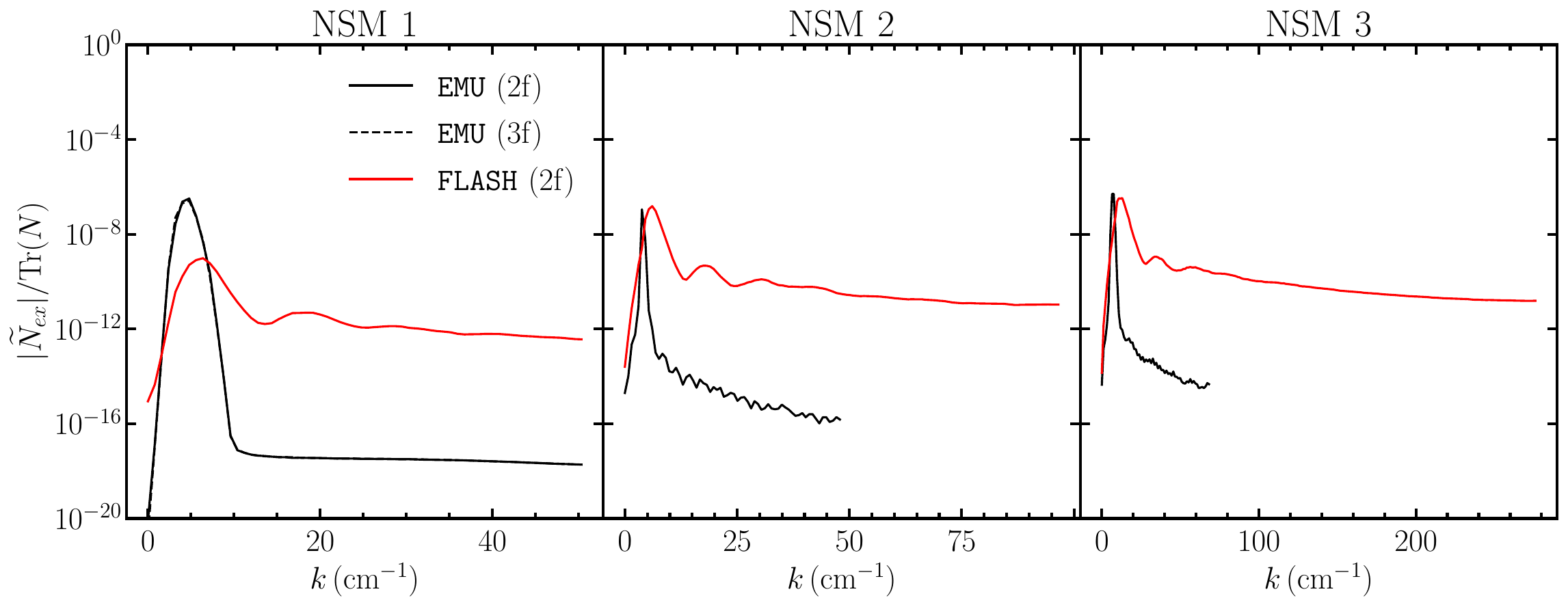}
    \caption{The magnitude of the discrete Fourier transform of $N_{ex}$ in the three NSM simulations plotted against wavenumber.  DFTs are at a time prior to saturation.
    For NSM 1, both \flash and \emu DFTs are $0.1\,{\rm ns}$ before saturation.
    For NSM 2, the DFT for \flash occurs $0.5\,{\rm ns}$ before saturation and $0.1\,{\rm ns}$ for \emu.  For NSM 3, the DFT for \flash occurs $0.15\,{\rm ns}$ before saturation and $0.05\,{\rm ns}$ for \emu.
    }
    \label{fig:FFT_NSM}
\end{figure*}

\paragraph{Fourier space analysis} Figure \ref{fig:FFT_NSM} gives the DFTs for the three NSM points at $0.1\,{\rm ns}$ before saturation, roughly corresponding to the second panel in Fig.\ \ref{fig:volume_rendering}.  Because the growth rates are different for the second (third) points, we pick a time before saturation of $0.5\,{\rm ns}$ ($0.15\,{\rm ns}$) for \flash, and $0.1\,{\rm ns}$ ($0.05\,{\rm ns}$) for \emu, all of which allow us to most clearly capture the fastest growing mode before it begins to non-linearly couple to other modes.
For the first NSM point, we find $\kmax=6.4\,{\rm cm}^{-1}$, corresponding to a wavelength of $1.0\,{\rm cm}$ and matching the distance between planar structures in the second panel of Fig.\ \ref{fig:volume_rendering} (domain size $L\sim8\,{\rm cm}$).

Similar to the results of Sec.\ \ref{sec:results_tests}, the DFTs from the \flash simulations show fastest growing modes with smaller wavelengths, higher noise floors, and visible harmonics for all three simulation points. The \emu simulations were run with a larger grid cell size to most optimally resolve the larger unstable wavelengths, resulting in a DFT that is cut off at smaller maximum $k$. Despite the differences between the \flash and \emu calculations, we emphasize that all of the \flash simulations once again reflect characteristic FFI behavior with discernible peaks with reasonable values of the fastest growing mode.

\begin{table*}
    \centering
    \hspace{-1.5cm}
    \begin{tabular}{c|ccccc}
        \multirow{2}{*}{Name} &  \multirow{2}{*}{$\langle|N_{ex}|/N_{ee}(0)\rangle|_{t=t_{\rm sat}}$} & \multirow{2}{*}{$\langle|N_{ex}(t_{\rm dec})|\rangle/\langle|N_{ex}(t_{\rm sat})|\rangle$} & \multirow{2}{*}{$\langle N_{ee}/N_{ee}(0)\rangle|_{t\rightarrow\infty}$} & \imo & \kmax \\
         &&&& $(10^{10}\,\mathrm{s}^{-1})$ & $(\mathrm{cm}^{-1})$ \\\hline
        NSM 1 & & & & & \\
        \flash     & $0.158$ & $0.0766$ & $0.643$ &
        $8.1$ &
        $6.4(4)$\\
        \emu (2f)  & $0.178$ & $0.306$ & $0.743$ &
        $5.6$ &
        $4.8(4)$\\ \hline
        NSM 2 & & & & & \\
        \flash     & $0.0845$ & $0.0395$ & $0.837$ &
        $5.2$ &
        $6.1(4)$\\
        \emu (2f)  & $0.0640$ & $0.413$ & $0.960$ &
        $1.1$ &
        $3.8(4)$\\ \hline
        NSM 3 & & & & & \\
        \flash     & $0.181$ & $0.00707$ & $0.609$ & $10.7$ &
        $13.0(5)$\\
        \emu (2f)  & $0.170$ & $0.624$ & $0.831$ &
        $4.2$ &
        $6.5(5)$\\ \hline
    \end{tabular}
    \caption{Results for \flash and \emu (2f) calculations for the three 3D NSM simulations. Column labels are the same as Table \ref{tab:3test_results}.
    }
    \label{tab:nsm_results}
\end{table*}

To summarize, Table \ref{tab:nsm_results} gives numerical results of FFC to compare between \flash and the 2-flavor \emu calculations for the three NSM simulations.
The results in Table \ref{tab:nsm_results} are presented in the same way as Table \ref{tab:3test_results}.
Like the Fiducial, 90Degree, and TwoThirds tests, the \flash calculations for the three NSM points have larger values of \imo and \kmax compared to \emu, but still exhibit reasonable behavior characteristic of the FFI.

\section{Conclusions}
\label{sec:conclusion}

Core collapse supernovae and merging neutron stars are complex systems that require the melding of many different physics aspects including magnetohydrodynamics, general relativity, equation of state physics, and neutrino physics.   When neutrino moment methods are used currently in large scale simulations, they consist of classical neutrino physics and typically employ two angular moments with a closure for each neutrino species.  In this work, we have extended the moment method framework in the context of the \flash code to take into account neutrino flavor transformation in a two-moment scheme with a quantum closure.  While at present we have tested the flavor transformation alone, the success of the moment method in modeling many features of the fast flavor instability lends confidence to incorporating it into large-scale multi-physics simulations in the future.

We found that neutrino transformation behavior is well captured in a number of test problems, including vacuum oscillations, Mikheyev-Smirnov-Wolfenstein (MSW) resonance, and bipolar oscillations.  Additionally, we approximately reproduced the results of three multidimensional PIC simulations in the literature at a fraction of the computational cost.  For those tests, in the key metric of the final electron neutrino number density, we find very good agreement among the first two ($\sim1\%$) and more qualitative agreement with the last ($\sim15\%$).
Similarly, the two-moment simulations were able to almost exactly match instability growth rates in the first two although our moment method has somewhat faster growth in the last.
The moment method also shows fastest growing modes that peak at a similar, but slightly higher wave number than in the PIC calculations. Further analysis is required to tease out the details of the dispersion relation under the two-moment approximation in regions of instability \citep{Froustey:2023skf} and compare with numerical simulation. 

We then performed multidimensional simulations of the FFI in three separate neutrino angular distributions taken from a full-scale classical simulation of an NSM.  In key quantities, we found similar levels of agreement as we found in the TwoThirds test. 
Specific differences include a lower electron neutrino number density at saturation with the moment method than with \emu, and a faster growth rate with the moment method.
While we always found qualitative agreement, the different methods naturally show the largest deviations when the distribution is only marginally unstable. Quantitative agreement is best with deep crossings and tends to worsen in the case of shallow crossings. This shortcoming will be important to improve upon in future advances of the algorithm, since the rapid onset of the FFI is likely to drive ELN crossings to remain shallow in astrophysical environments.

Our two-moment algorithm is based on an extension of the classical MEC relevant to quantum neutrino transport \citep{PhysRevD.102.083017}.
Figures \ref{fig:pressure_fid} and \ref{fig:pressure_nsm1} show that the classical MEC by itself cannot fully characterize the underlying distribution, and most certainly leads to discrepant results for the pressure tensor on the order of a few percent.
As the closure accounts for missing physics from the unevolved higher-order moments, we anticipate that future efforts to develop quantum closures will improve the agreement between this two-moment method and more exact methods. To go beyond the simple prescription implemented here requires using all of the components of $E$ and $\mathbf{F}$ in a basis-independent way as suggested by \cite{kneller23}.

Because of the small scales on which flavor transformation occurs, our method is at present still too computationally expensive to directly place in a full-scale CCSN or NSM simulation.  However, we anticipate that in the future, methods can be developed to incorporate the very small-scale physics of flavor transformations into large-scale simulations.  Nevertheless, by virtue of only following two moments, this method is substantially computationally cheaper than exact ones that evolve neutrino distributions along hundreds or thousands of directions. Given the successes of capturing many features of the FFI in this work, we believe that moment methods can complement the higher fidelity methods such as PIC and multi-angle.  For example, moment methods could be useful in doing faster realistic calculations when searching configuration space \citep{Johns:2021taz}, with follow-up post-processing being done by PIC or multi-angle codes.
Machine learning is also a possibility, with moments being used to train the algorithms \citep{Abbar:2023kta}.

Despite the caveats above, we find that our angular-moment implementation of the QKEs reasonably and effectively captures the complex and confounding phenomenon of neutrino flavor transformation in conditions which are plausible in the environments of CCSNe and binary NSMs.  Specifically, in anisotropic conditions where the angular neutrino distributions exhibit a lepton number crossing, 
the two moments of an M1 transport scheme manifest the phases of fast flavor conversion: exponential growth during unstable conditions; peak saturation of the off-diagonal density matrix component; rapid flavor oscillations of the diagonal components; and post-saturation decoherence with subsequent freeze-out.  

There are many possible extensions that one can make to the work presented here.  Improvements on the closure, the addition of the collision matrix, and an extension to three flavors are a few. Finally, while we have included advection in our simulations, the inclusion of both advection and flavor mixing in the context of a large scale simulation will likely alter the angular distributions of the neutrinos \citep{Padilla-Gay:2020uxa,2023arXiv230610108N}. These improvements will widen the conditions for which this method can be used and allow us to probe other predicted phenomena, such as collisional instabilities, bipolar oscillations and matter neutrino resonance transitions.

Including flavor transformation in 3D general-relativistic-magnetohydrodynamics astrophysics simulations is a major computational challenge for multi-messenger-astrophysics theory.
Our moment-method flavor calculations offer a contribution to this important field of study.

\section{Acknowledgments}

We thank George Fuller and MacKenzie Warren for useful conversations.
EG, JPK, and GCM are supported by the Department of Energy Office of Nuclear Physics award DE-FG02-02ER41216. EG acknowledges support by the National Science Foundation grant  No.\ PHY-1430152 (Joint Institute for Nuclear Astrophysics Center for the Evolution of the Elements). 
SMC is supported by the U.S. Department of Energy, Office of Science, Office of Nuclear Physics, Early Career Research Program under Award Number DE-SC0015904. JF is supported by the Network for Neutrinos, Nuclear Astrophysics and Symmetries (N3AS), through the National Science Foundation Physics Frontier Center Grant No. PHY-2020275. SR is supported by a National Science Foundation Astronomy and Astrophysics Postdoctoral Fellowship under award AST-2001760. FF is supported by the Department of Energy, Office of Science, Office of Nuclear Physics, under contract number
DE-AC02-05CH11231, by NASA through grant
80NSSC22K0719, and by the National Science Foundation through grant AST-2107932.
The calculations presented in this paper were undertaken, in part, on the \emph{Payne} machine at North Carolina State University which is supported in part by the Research Corporation for Science Advancement. This research used resources of the National Energy Research Scientific Computing Center (NERSC), a U.S. Department of Energy Office of Science User Facility located at Lawrence Berkeley National Laboratory, operated under Contract No. DE-AC02-05CH11231 using NERSC award ERCAP0021631.  This material is based upon work supported by the U.S. Department of Energy, Office of Science, Office of Advanced Scientific Computing Research and Office of Nuclear Physics, Scientific Discovery through Advanced Computing program under Award Number DE-SC0017955.
The authors would like to acknowledge the use of the following software: Matplotlib \citep{hunter:2007}, Numpy \citep{vanderwalt:2011}, and SciPy \citep{scipy}.

\appendix

\section{1D Test Problems}
\label{app:test_problems}

To test the algorithms for simultaneous flavor-transformation and advection we implemented in \flash, we have designed and conducted three tests in 1D to quantify the numerical accuracy and precision of the code. This appendix contains the results of those tests, which we term: Vacuum; MSW; and Bipolar Oscillations. We compare to analytic solutions for the vacuum and MSW tests, and to an analytic prediction for the period of the bipolar test.

\subsection{Vacuum Oscillations}

In this test problem we consider only the vacuum potential $H_V$ in the QKEs of Eqs.\ \eqref{eq:qke_E} and \eqref{eq:qke_F}, but with a transformation into spherical coordinates.  Neutrinos and anti-neutrinos are emitted isotropically from a point source at the origin. The domain has an outer radius of $4\,{\rm km}$. We use a spherical geometry with an inner radius of $r_0=100\,{\rm m}$ to avoid divide by zero errors at the origin, and assume the flux factors at $r_0$ are all unity as the neutrinos stream outwards in the radial direction. The number of grid points is 640, all evenly-spaced in radius.  Initially, all cells have the same density of the four particle species scaled by $1/r^2$.  The flux factors are set to 1.0.  The energy densities of the off-diagonal species are set to zero.  We have set the outer boundary condition to an outflow, and the inner boundary to that of an oven emitting $\nu_e,\overline{\nu}_e,\nu_x,$ and $\overline{\nu}_x$.  At time $t=0$ we switch on the emission of neutrinos and antineutrinos at a constant rate from a source located at $r_0$ moving in the positive outward direction. The neutrinos are $99\%$ the electron flavor and $1\%$ $x$-flavor; the antineutrinos are $90\%$ the electron antineutrino flavor and $10\%$ the $x$-antineutrino flavor. For $r < c\,t$ the solution of the moment transport equations is that 
\begin{align}
  \frac{E_{ee}^{\rm (th)}(r)}{{\rm Tr}[E^{\rm (th)}(r)]} &= [1-P^{(V)}_T(r)]\frac{E_{ee}(r_0)}{{\rm Tr}[E(r_0)]} + P^{(V)}_T(r)\frac{E_{xx}(r_0)}{{\rm Tr}[E(r_0)]}
  \label{eq:vac_pred_ee}\\
  \frac{E_{xx}^{\rm (th)}(r)}{{\rm Tr}[E^{\rm (th)}(r)]} &= [1-P^{(V)}_T(r)]\frac{E_{xx}(r_0)}{{\rm Tr}[E(r_0)]} + P^{(V)}_T(r)\frac{E_{ee}(r_0)}{{\rm Tr}[E(r_0)]}
  \label{eq:vac_pred_xx}
\end{align}
where the function $P^{(V)}_T$ is the flavor transition probability 
\begin{align}
P^{(V)}_T(r) &= \sin^2(2\theta)\,\sin^2\left[\frac{\delta m^2}{4p}(r-r_0)\right],
\end{align}
$\theta$ is the mixing angle, $\delta m^2$ is the squared neutrino mass difference, and $p$ is the neutrino energy.
Analogous expressions to Eqs.\ \eqref{eq:vac_pred_ee} and \eqref{eq:vac_pred_xx} for the theoretical anti-neutrino energy density moments exist with an identical transition probability $\overline{P}^{(V)}_T(r)=P^{(V)}_T(r)$.

\begin{figure}[htb!]
    \centering
    \includegraphics[scale=0.45]{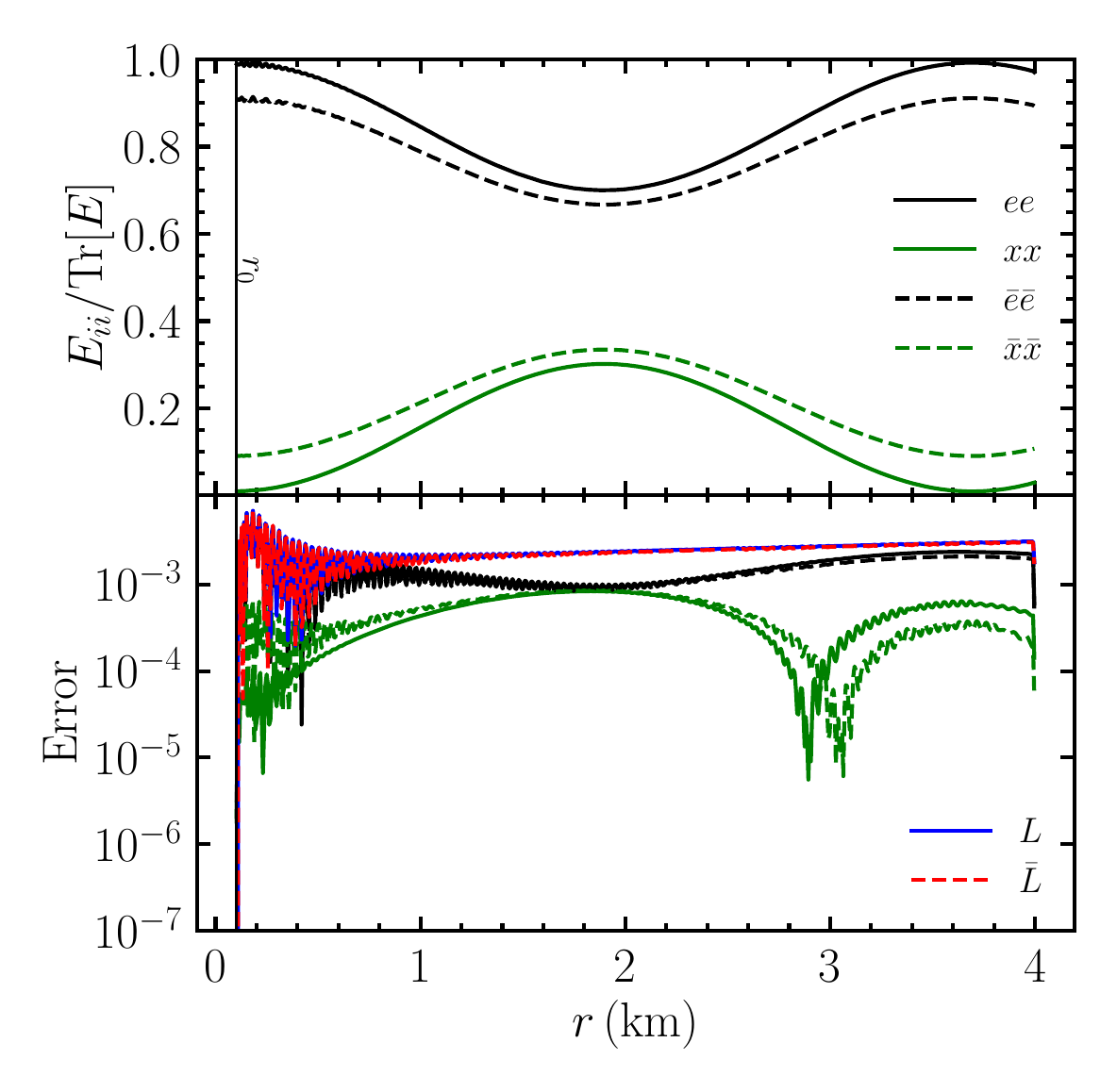}
    \caption{Results from the vacuum test of \flash plotted against radial coordinate.  [Top] Diagonal density matrix components of the energy density moment ($E_{ii}$) normalized by the trace of $E$.  Barred components denote the anti-neutrino counterparts, i.e., $E_{\bar{e}\bar{e}}\equiv\overline{E}_{ee}$ and are scaled by the trace of $\overline{E}$.  [Bottom] Errors in various components.  Green and black lines correspond to difference between the calculated $E_{ii}$ and the theoretical predictions for Eqs.\ \eqref{eq:vac_pred_ee} and \eqref{eq:vac_pred_xx} scaled by the trace of $E$.  Anti-neutrino counterparts follow from similar equations.  Solid blue (dashed red) correspond to the difference between the calculated length of the polarization vector and the initial length of the polarization vector for neutrinos (anti-neutrinos) scaled by the initial polarization vector length.}
    \label{fig:vacuum_test}
\end{figure}

Figure \ref{fig:vacuum_test} shows the results of this test at a time $t=3.3\times10^{-5}\,{\rm s}$. We use a mixing angle $\theta=0.28818$, a mass-squared difference $\delta m^2=6.9\times10^{-4}\,{\rm eV}^2$, and an energy $p=1\,{\rm MeV}$.  The top panel of the figure shows the flavor content evolving with the familiar oscillatory pattern. We use the shorthand notation of $E_{\bar{e}\bar{e}}\equiv\overline{E}_{ee}$ for the electron anti-neutrino energy density moment, and similarly for the $\bar{x}\bar{x}$ component.  In the lower panel we plot the relative error of the numerical solution compared to the analytic results in Eqs.\ \eqref{eq:vac_pred_ee}, \eqref{eq:vac_pred_xx}, and the anti-neutrino analogs, namely
\begin{equation}\label{eq:rel_error}
  {\rm Error} = \frac{|Q-Q^{{\rm (th)}}|}{Q^{\rm (th)}},
\end{equation}
for the relevant quantity $Q$.  In addition to the errors in the energy density moments, we give the relative error in the Bloch polarization vector for neutrinos, $L$, and anti-neutrinos, $\overline{L}$.
The dominant source of the error arises from the finite grid spacing in radius.  We have verified that decreasing/increasing the number of grid points increases/reduces the errors. The size of the error scales inversely with the number of grid points.
Notice that the largest errors occur at small radii for $E_{ee}$ and $\overline{E}_{ee}$ (and consequently also for $L$ and $\overline{L}$) at a level of a few parts in $10^3$.  This error is due to numerical oscillations at small radii which are a result of the inner boundary condition in spherical coordinates and not related to flavor-transformation.  Overall, the \flash calculation is able to successfully reproduce the analytic results to a precision better than $\sim 1\%$. 

\subsection{MSW Oscillations}

Our second 1D test case is similar to the vacuum test case described in the previous section.  In addition to the vacuum potential, we now add a constant matter density to observe MSW oscillations \citep{1986PhRvL..56.1305B,1986PhRvL..57.1271H}.
Our domain ranges from the inner boundary at $r_0=10\,{\rm km}$ to the outer boundary at $r=20\,{\rm km}$.  We use 320 cells evenly-spaced in the radial coordinate.  The evolution equations for this test can actually be transformed such that the analytic solution to the energy density moments is the same as Eqs.\ \eqref{eq:vac_pred_ee} and \eqref{eq:vac_pred_xx} with a different MSW transition probability
\begin{equation}\label{eq:pt_msw}
P^{(M)}_T(r) = \sin^2(2\theta_{\rm eff})\sin^2\left[\frac{\delta m^2_{\rm eff}}{4p}(r-r_0)\right],
\end{equation}
\begin{equation}
\sin^2(2\theta_{\rm eff}) = \frac{\sin^2(2\theta)}{\sin^2(2\theta)+C^2},
\end{equation}
\begin{equation}
\delta m^2_{\rm eff} = \delta m^2\sqrt{\sin^2(2\theta)+C^2},
\end{equation}
\begin{equation}\label{eq:c_msw}
C = \cos(2\theta) - \frac{2\sqrt{2}\,p\,G_FY_e\rho}{\delta m^2\,m_u},
\end{equation}
where $\rho$ is the matter density, $Y_e$ the electron fraction, and $m_u$ the atomic mass unit.
For anti-neutrinos $\overline{P}^{(M)}_T$ is calculated the same as above but with the placement of a minus sign on the electron fraction, $Y_e\rightarrow -Y_e$. We can then compare this solution to the \flash calculation, which simply adds $H_M$ to $H_V$ in Eqs.~\eqref{eq:qke_E} and \eqref{eq:qke_F}.

\begin{figure}[htb!]
    \centering
    \includegraphics[scale=0.45]{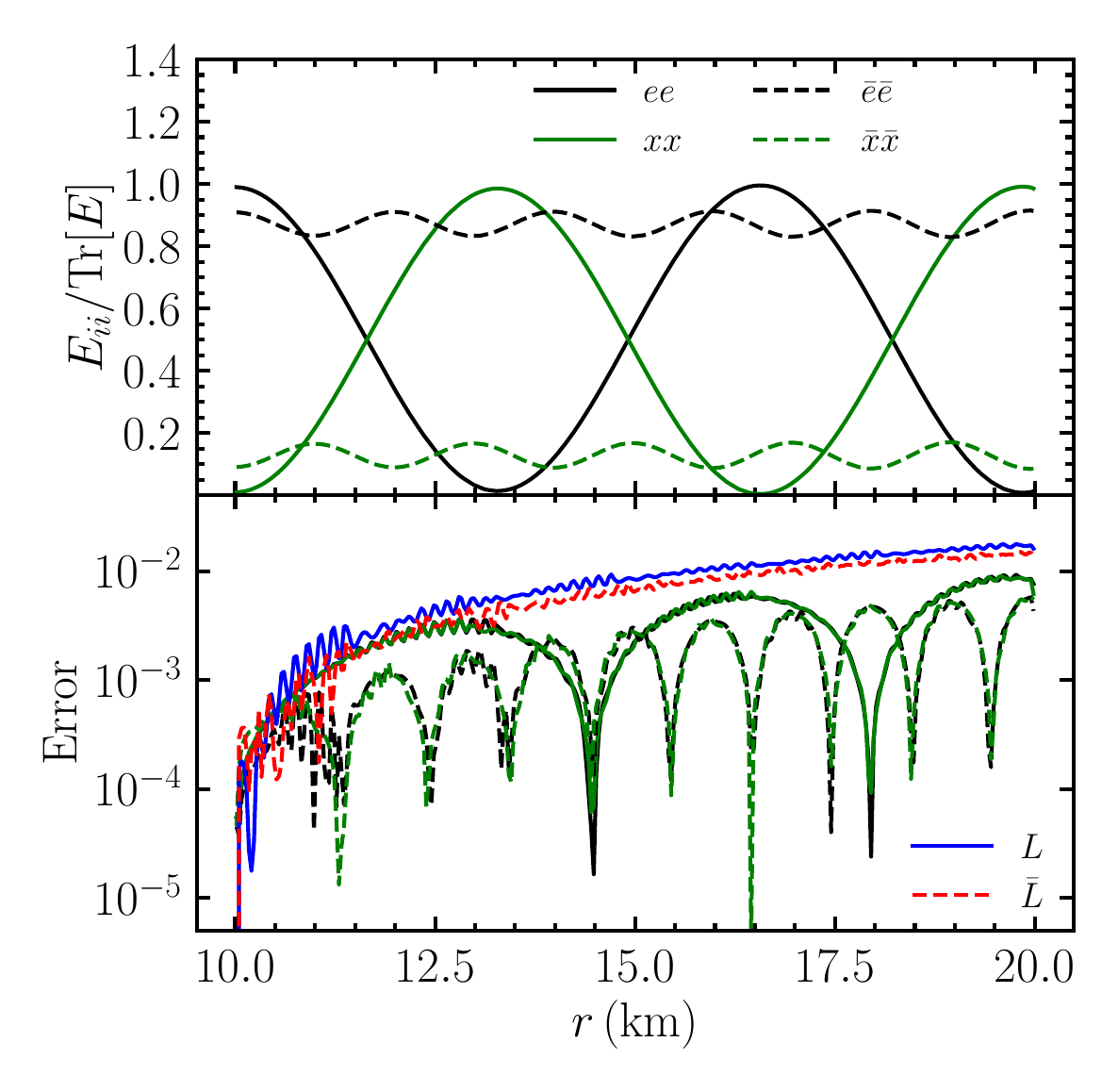}
    \caption{Results from the MSW test of \flash plotted against radial coordinate.  Notation for both panels is the same as Fig.\ \ref{fig:vacuum_test}. Error for the $E_{ii}$ components corresponds to the difference between the calculated $E_{ii}$ and the theoretical predictions using the transition probability in Eq.\ \eqref{eq:pt_msw} for neutrinos, and a similar expression for anti-neutrinos.}
    \label{fig:msw_test}
\end{figure}

The results of this test case at $t=1.67\times10^{-4}\,{\rm s}$ are shown in Fig.\ \ref{fig:msw_test}, where the panels have the same plotting conventions as those in Fig.\ \ref{fig:vacuum_test}. We use a matter density of $\rho=8\times10^3\,\mathrm{g\,cm}^{-3}$ and an electron fraction $Y_e=0.5$. The familiar oscillatory evolution of the neutrinos and anti-neutrinos as a function of distance is apparent. The difference in period between the neutrinos and anti-neutrinos is due to the replacement of $Y_e$ by $-Y_e$ in Eq.\ \eqref{eq:c_msw}.  The errors for this MSW test case are of a similar order of magnitude as the vacuum case.

\subsection{Bipolar Oscillations}
\label{app:bipolar}

Our final test case is a study of bipolar neutrino oscillations. 
This system isolates and tests the numerical convergence in our flavor-mixing subroutines by obviating the advection processes.  When using negative energy eigenvalues for the anti-neutrino density matrices, systems that are represented by two nearly oppositely directed flavor polarization vectors are termed \emph{bipolar systems} \citep{2010ARNPS..60..569D}.
As they evolve, the neutrino polarization vector nearly flips in direction, while the anti-neutrino polarization vector exhibits the same behavior but $\pi$ radians out of phase. In our treatment, we use positive energy eigenvalues for the anti-neutrinos in Eq.\ \eqref{eq:bar_H}, and so we begin with aligned flavor polarization vectors to study the bipolar behavior. As a result, we would expect bipolar oscillations in phase between the neutrinos and anti-neutrinos.  We study a simple case where the system is homogeneous and monochromatic. Only the $H_V$ and $H_\nu$ terms are included in the Hamiltonian for Eqs.\ \eqref{eq:qke_E} and \eqref{eq:qke_Ebar}.  The flux moments are all zero. We pick $\delta m^2/p$ such that the period of vacuum oscillations is $2\pi\, {\rm s}$, and take a small vacuum mixing angle $\theta = 0.01$.  The energy density is picked such that the strength of self-interactions is $10\,{\rm rad/s}$.

\begin{figure}[htb!]
    \centering
    \includegraphics[width=0.5\textwidth]{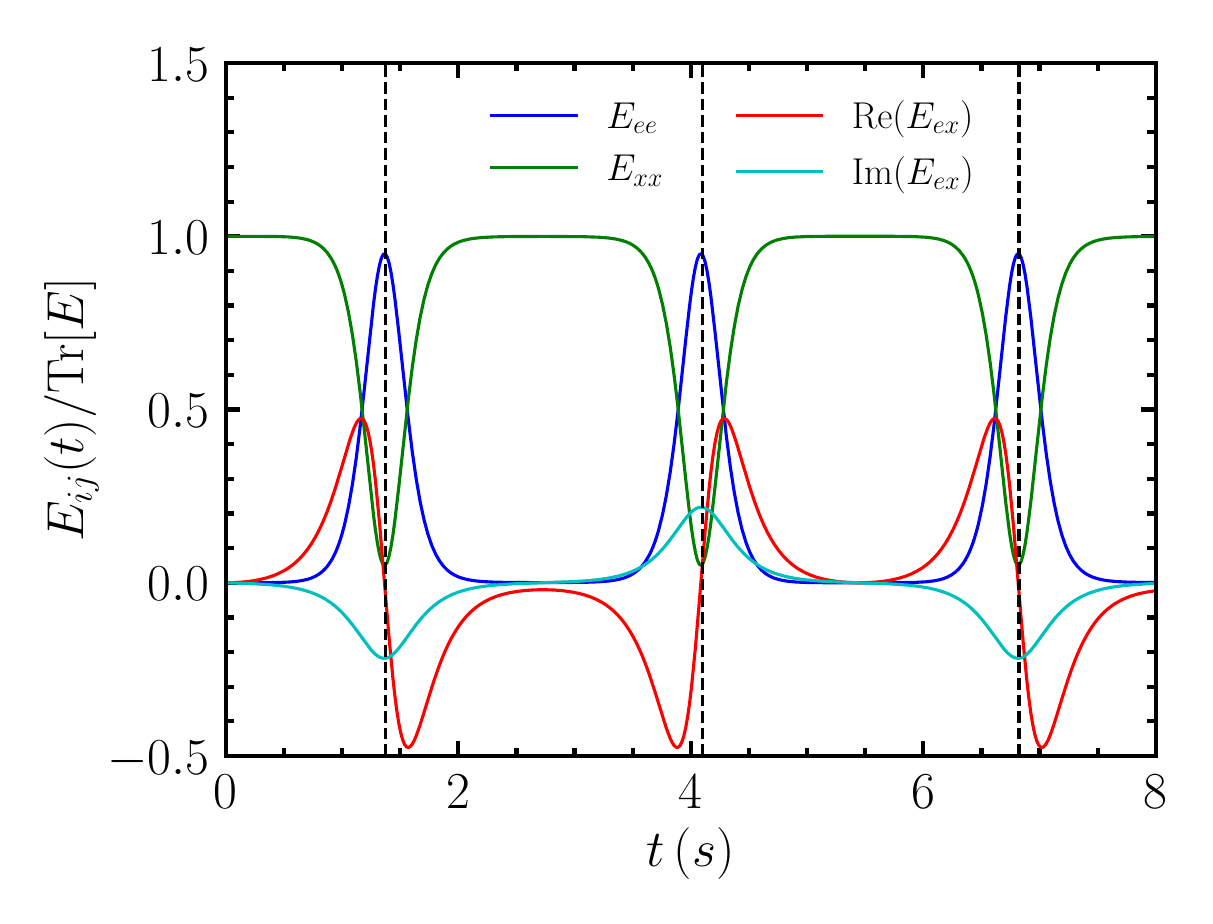}
    \caption{Results from the bipolar test of \flash plotted against time.  Only shown are density matrix components for neutrinos.
    We include vertical dashed black lines to indicate the period of the oscillations based on the analytical result of Ref.\ \cite{PhysRevD.74.105010}.}
    \label{fig:bipolar_test}
\end{figure}

Figure \ref{fig:bipolar_test} shows the results of our bipolar test for neutrinos — the anti-neutrinos evolve in the same manner.
The various color curves label the individual components of the energy density matrix.
Initially the neutrinos are completely in $x$-flavor eigenstates. They subsequently flip to a state which is $\sim95\%$ electron flavor, before reverting back to the initial state.  This pattern repeats indefinitely as the system is always unstable.
Vertical dashed black lines give predictions for the locations in time of the period of oscillations in $E_{xx}$, centered on the first peak of the $E_{ee}$ curve. We use a closed-form expression involving elliptic integrals to predict the period of oscillation in the unstable bipolar system.  Note that the period for the flavor off-diagonal components, ${\rm Re}(E_{ex})$ and ${\rm Im}(E_{ex})$ have double the period as the diagonal components. There exists a slight decrease in the period of the \flash results, which becomes more apparent as time continues.  At the first full period near $t\sim4.1\,{\rm s}$, the relative difference in expected versus calculated time is  $1.7\times10^{-3}$.  At the second full period, near $t\sim6.8\,{\rm s}$, the drift is $2.1\times10^{-3}$.  Both of these drift values are within the size of the time step.

\begin{figure}[htb!]
    \centering
    \includegraphics[width=0.5\textwidth]{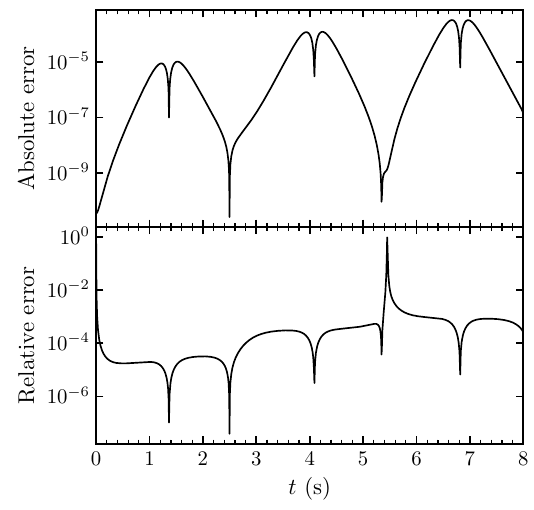}
    \caption{Relative and absolute errors for $E_{ee}(t)$ between the \flash calculation and a direct numerical resolution of Eq.~(12) in~\cite{PhysRevD.74.105010}}.
    \label{fig:bipolar_error}
\end{figure}

We compare this calculation using the \flash architecture to a more direct and straightforward resolution of the equation of motion\footnote{For the interested reader: \citet{2023PhRvD.108d3007X} gives an analytical formulation of the curves in Fig.\ \ref{fig:bipolar_test}.}, based on the study of~\cite{PhysRevD.74.105010}. Using the \texttt{solve{\_}ivp} function from the Scipy library with high precision parameters ($\mathtt{rtol}=\mathtt{atol}=10^{-8}$), we solve Eq.~(12) from \citet{PhysRevD.74.105010} for the tilt angle $\varphi$.  We can solve for the trace-normalized value of $E_{ee}$ by taking the tilt angle and solving for the Bloch polarization vector [see Eqs.~(6) and (8) of \citet{PhysRevD.74.105010}].\footnote{Since we consider a normal hierarchy scenario with an initial system of $x$ (anti)neutrinos instead of electronic ones, there are some minor changes to the equations in~\citep{PhysRevD.74.105010}. Borrowing their notations:
\[\begin{aligned}
    (8) \ \to \ Q &= \left[4 + \left(\frac{\omega}{\mu}\right)^2 - 4 \frac{\omega}{\mu} \cos 2 \theta_0\right]^{1/2} \, ,  \\
    \varphi(0) &\simeq \pi + (\omega/\mu Q)2\theta_0 \, , 
\end{aligned}\] with here $\mu/\omega = 10$ and $\theta_0 = 0. 01$.} Figure \ref{fig:bipolar_error} gives the absolute (top panel) and relative (bottom panel) errors between \flash and the Scipy function \texttt{solve\_ivp} for the quantity $E_{ee}/{\rm Tr}[E]$.  The relative error is the same expression as Eq.\ \eqref{eq:rel_error}, but we substitute the results from \texttt{solve\_ivp} for $Q^{\rm (th)}$.  The absolute error is the numerator of the rhs of Eq.\ \eqref{eq:rel_error}.  We see an overall growth in the peak absolute error with increasing cycle number, consistently with the drift in the period of oscillations.  Both the absolute and relative errors are the result of numerical convergence in our RKCK solver.  We have verified that these errors scale linearly with the tolerance criterion of the RKCK algorithm.

\section{Moments in an Orthonormal Tetrad}
\label{sec:ortho}

In order to perform simulations of realistic neutrino distributions in a small domain using codes that assume a Minkowski metric, we need to transform the radiation field quantities output by SpEC-Hydro into an orthonormal tetrad comoving with the background fluid. The radiation field is given in terms of the lab-frame energy density $E^{(\nu_a)}$, energy flux $F_i^{(\nu_a)}$, the average neutrino energy $\epsilon^{(\nu_a)}$, and the fluid transport velocity $\widetilde{v}^i$ at each point in space and for each neutrino flavor $a$. The spacetime metric is given in standard 3+1 quantities: the lapse $\alpha$, shift $\beta^i$, and 3-metric $\gamma_{ij}$. This is a common procedure within general-relativistic truncated moment simulation codes \citep{10.1143/PTP.125.1255, FoucartM1:2016, Kuroda_2016}, but we make the procedure explicit here for completeness. Note that in the main text, we revert to the more conventional symbols $E$ and $\mathbf{F}$ to indicate tetrad quantities, but they refer to the lab-frame quantities in this appendix for consistency with \cite{10.1143/PTP.125.1255}.

The radiation stress-energy tensor can be defined in either the lab or comoving frames as \citep{10.1143/PTP.125.1255}
\begin{equation}
\begin{aligned}
    T^{\mu\nu}&= E n^\mu n^\nu + F^\mu n^\nu + n^\mu F^\nu + P^{\mu\nu} \,\, \mathrm{[lab]} \\
    &= J u^\mu u^\nu + H^\mu u^\nu + u^\mu H^\nu + S^{\mu\nu}\,\,\mathrm{[comoving]}
\end{aligned}
\label{eq:stress_energy}
\end{equation}
To get the moments in the comoving frame, we evaluate the full stress-energy tensor and project onto suitably chosen tetrad basis vectors. This is made more complicated by the fact that the pressure tensor needs to be expressed as a linear combination of the optically thick and thin limits according to the two-moment closure.

The index on the shift vector can be lowered with the three-metric ($\beta_i = \beta^j \gamma_{ij}$). The vector that is normal to the hypersurface of constant time $t$ can be constructed as $n_\mu=\alpha(-1,0,0,0)$. The transport velocity is defined in terms of the four-velocity as $\widetilde{v}^i=u^i/u^t$, where $W=\alpha u^t$ is the generalized Lorentz factor. From these it is straightforward to construct the fluid four-velocity $u^\alpha$ and the three-velocity $v^i=u^i/W + \beta^i/\alpha$.

The optically thin pressure tensor is straightforwardly defined as $P^{ij}_\mathrm{thin}=(E/F^k F_k) F^i F^j$, where contractions are naturally evaluated using the lab-frame three-metric $\gamma_{ij}$. 
Obtaining $P^{ij}_\mathrm{thick}$ is more involved, but can be determined analytically as follows. It is straightforward to write the optically thick pressure tensor in the lab frame as $S^{ij}_\mathrm{thick}=Jh^{ij}/3$, where $h^{ij}=g^{ij}+u^i u^j$ is the comoving-frame three metric. By contracting Eq.~\eqref{eq:stress_energy} with various combinations of $u^\alpha$ and $n^\alpha$, one can work out that the comoving-frame radiation energy density is [i.e., starting with Eq.\ (87) in \citet{FoucartM1:2016}]
\begin{equation}
\begin{aligned}
    J &= T^{\mu\nu} u_\mu u_\nu \\
    &=W^2\left[E-2v^i F_i + J W^2 (v_i v^i)^2 + 2 W H_i v^i (v_i v^i) + S^{ij}v_i v_j\right]
    \end{aligned}
\end{equation}
Plugging in $S^{ij}$ from above, this in turn yields the comoving-frame energy density in the optically thick limit:
\begin{equation}
    J_\mathrm{thick} = \frac{3}{2 W^2+1}\left[(2W^2-1)E-2W^2 v^i F_i\right]
\end{equation}
Similarly,
\begin{equation}
    \begin{aligned}
    H^i_\mathrm{thick} &= -T^{\mu\nu}u_\mu (g_\nu^{\phantom{\nu}i} + u_\nu u^i)\\
    &= \frac{F^i}{W} + \frac{v^i W}{2W^2+1}\left[(4W^2+1)F_iv^i - 4W^2 E\right] 
    \end{aligned}
\end{equation}
Finally, these can be used to calculate the pressure tensor in the optically-thick limit as
\begin{equation}
    \begin{aligned}
    P^{ij}_\mathrm{thick} &= T^{\mu\nu}(g_\mu^{\phantom{\mu}i}+u_\mu u^i)(g_\mu^{\phantom{\nu}j}+u_\mu u^j)\\
    &= \frac{J_\mathrm{thick}}{3}(4 W^2 v^i v^j + \gamma^{ij}) + W(H^i_\mathrm{thick}v^j + v^i H^j_\mathrm{thick})
    \end{aligned}
\end{equation}
By plugging $E$, $F^i$, and $P^{ij}$ [from Eq.~\eqref{eq:Pij_closed}] into Eq.~\eqref{eq:stress_energy}, we now have the complete stress-energy tensor $T^{\mu\nu}$. 

For a comoving orthonormal tetrad, the timelike basis vector is the four velocity ($\hat{t}_\mu = u_\mu$). We then choose a spacelike trial vector $(0,1,0,0)$ (and similarly for the $y$ and $z$ trial vectors) and apply Gram-Schmidt orthonormalization to get the spacelike basis vectors $x^{(i)}_\mu$. The moments in the orthonormal tetrad are then
\begin{align}
  J_\mathrm{tet} &= T^{\mu\nu} \hat{t}_\mu \hat{t}_\nu,\label{eq:t_tet}\\
  H^i_\mathrm{tet} &= T^{\mu\nu} \hat{t}_\mu \hat{x}^{(i)}_\nu\label{eq:s_tet}.
\end{align}
From this point forward, we do not have to think about spacetime curvature, as at each location spacetime is locally flat. Specifically, the moments can now be used more intuitively. For example, the flux factor is simply $f=\sqrt{\sum_i (H^i_\mathrm{tet})^2}/J_\mathrm{tet}$.

\section{Differences in the initial conditions between \flash and \emu}
\label{app:diff}

\revision{
We must specify the initial conditions for all six of our test and NSM simulations shown in Sec.\ \ref{sec:results} across the entire domain.  Tables \ref{tab:3test_parameters} and \ref{tab:NSM_parameters} give the electron and $x$-flavor number densities and flux factors for the three tests and NSM points, respectively.
For the off-diagonal coherence terms, we use random numbers per the procedure outlined in Sec.\ \ref{subsec:initial_boundary_conditions}.  As a result of the randomization process, the complex valued $\varrho_{ex}$ terms add coherently which acts to reduce the domain-averaged values of the initial perturbation in $\langle |N_{ex}|\rangle$ for the \emu simulations. 
The first column of Table \ref{tab:app_table} gives $\langle|N_{ex}|/N_{ee}\rangle$ at the start of all six simulations for \flash and \emu (2f).
}

\begin{table*}
    \centering
    \hspace{-2.5cm}
    \begin{tabular}{c|cc}
        Name & $\langle|N_{ex}|/N_{ee}\rangle\vert_{t=0}$ & $t_{\rm sat}\,(10^{-9}\,{\rm s})$\\ \hline
        Fiducial & & \\
        \flash     & $7.65\times10^{-7}$ & $0.257$ \\
        \emu (2f)  & $9.85\times10^{-9}$ & $0.365$ \\ \hline
        90Degree & & \\
        \flash     & $7.65\times10^{-7}$ & $0.346$ \\
        \emu (2f)  & $9.85\times10^{-9}$ & $0.511$ \\ \hline
        TwoThirds & & \\
        \flash     & $7.65\times10^{-7}$ & $0.859$ \\
        \emu (2f)  & $8.53\times10^{-9}$ & $1.73$ \\ \hline
        NSM 1 & & \\
        \flash     & $7.65\times10^{-7}$ & $0.202$ \\
        \emu (2f)  & $1.19\times10^{-8}$ & $0.348$ \\ \hline
        NSM 2 & & \\
        \flash     & $7.65\times10^{-7}$ & $0.380$ \\
        \emu (2f)  & $6.19\times10^{-9}$ & $1.86$ \\ \hline
        NSM 3 & & \\
        \flash     & $7.65\times10^{-7}$ & $0.184$ \\
        \emu (2f)  & $1.42\times10^{-8}$ & $0.491$ \\ \hline
    \end{tabular}
    \caption{\revision{Initial values of domain-averaged off-diagonal magnitude of $N$ and saturation times for all six \flash and \emu (2f) FFC simulations.
    }}
    \label{tab:app_table}
\end{table*}

\begin{figure*}[htb!]
    \centering
    \includegraphics[width=\linewidth]{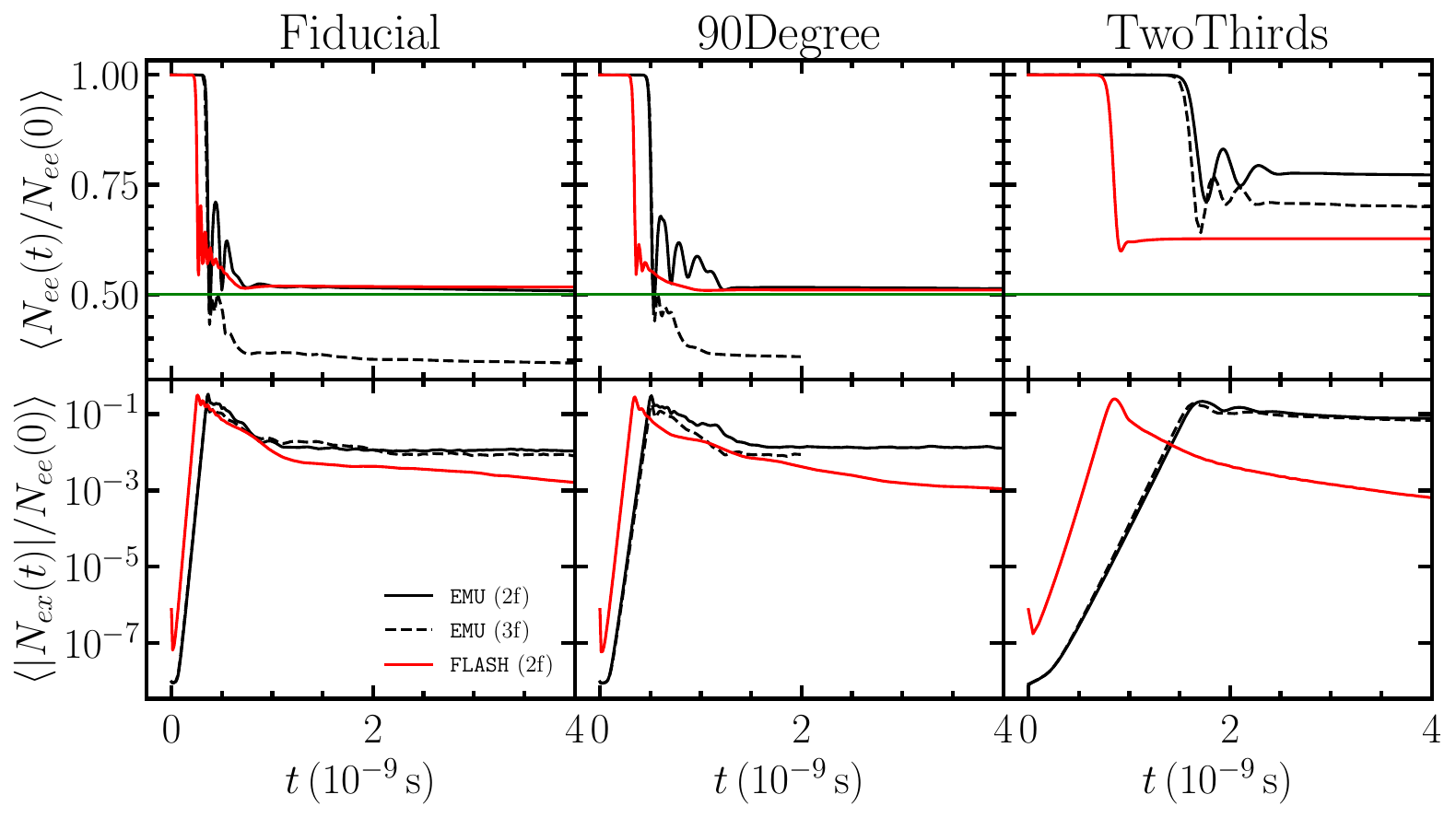}
    \caption{\revision{Density matrix elements versus time for the three tests in Sec.~\ref{sec:results_tests}. Figure content is the same as Fig.~\ref{fig:dm_tests} except all curves begin at simulation time $t = 0$.}
    }
    \label{fig:app_3test}
\end{figure*}

\begin{figure*}[htb!]
    \centering
    \includegraphics[width=\linewidth]{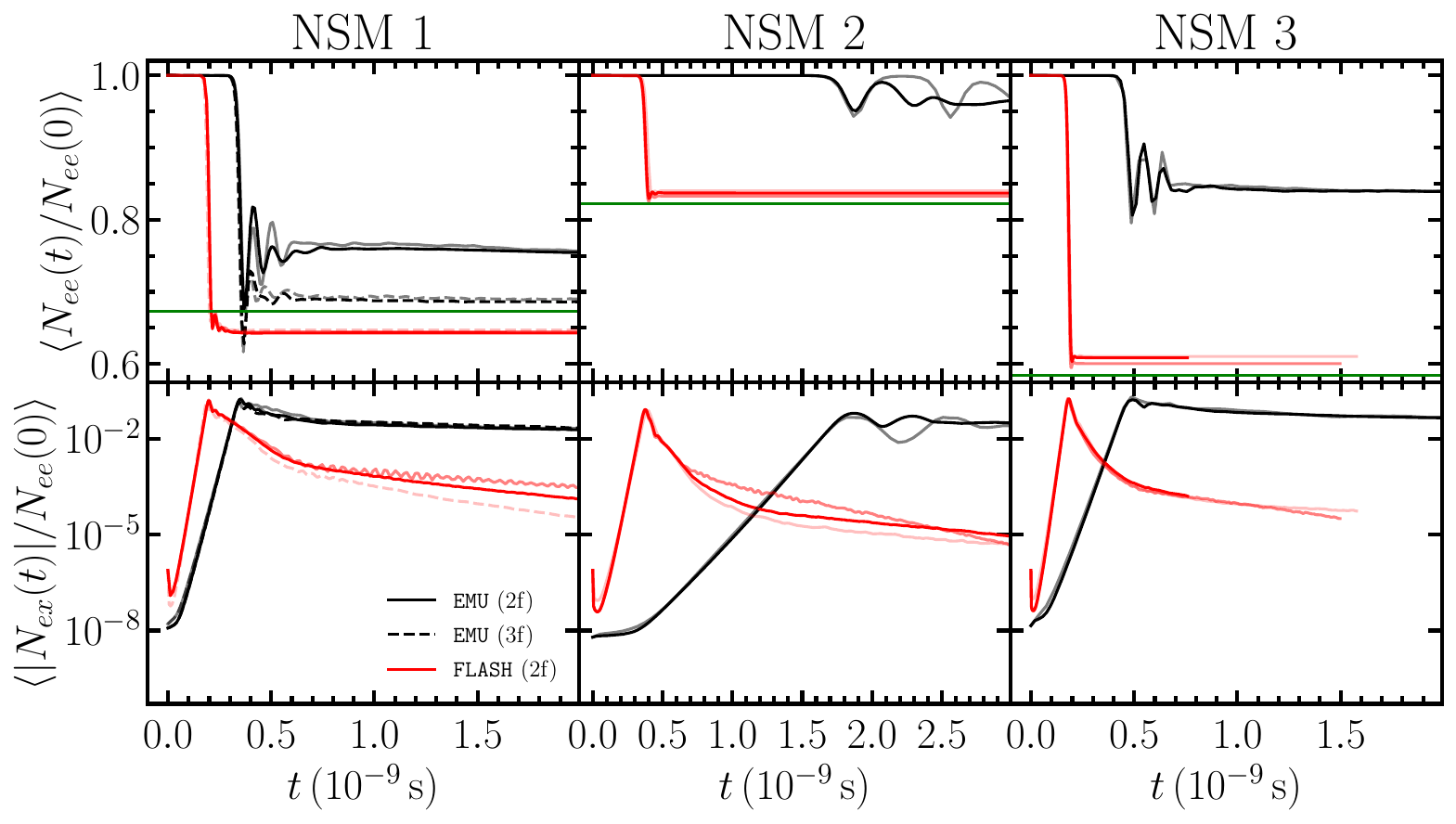}
    \caption{\revision{Density matrix elements versus time for the three NSM points in Sec.~\ref{sec:results_nsm}. 
    Figure content is the same as Fig.~\ref{fig:P_NSM} except all curves begin at simulation time $t = 0$.
    }}
    \label{fig:app_3NSM}
\end{figure*}

\revision{
Starting from smaller initial values, the \emu simulations will undergo different time evolution as compared to the \flash ones.
We give the evolution of $\langle N_{ee}\rangle$ and $\langle N_{ex}\rangle$ for the three tests of Sec.\ \ref{sec:results_tests} (Fig.\ \ref{fig:app_3test}) and the three NSM points of Sec.\ \ref{sec:results_nsm} (Fig.\ \ref{fig:app_3NSM}).  The content in Figs.\ \ref{fig:app_3test} and \ref{fig:app_3NSM} is the same as Figs.\ \ref{fig:dm_tests} and \ref{fig:P_NSM}, respectively, except for the definition of the horizontal axis, where we use simulation time for the figures in this appendix.  The smaller initial perturbations of $N_{ex}$ coupled with the smaller growth rates for \emu lead to later saturation points.  The second column of Table \ref{tab:app_table} gives the saturation times for all six simulations.
}

\bibliography{moment}

\end{document}

%% file: MFF_big_defs.tex
\newcommand{\flash}{{\tt FLASH}\xspace}
\newcommand{\emu}{{\tt EMU}\xspace}

\newcommand{\varrhobar}{\ensuremath{\overline{\varrho}}\xspace}

\newcommand{\ben}{\begin{enumerate}}
\newcommand{\een}{\end{enumerate}}